\documentclass[a4paper,11pt]{article}
\usepackage{graphicx}
\usepackage[abs]{overpic}
\usepackage{amsmath,amssymb}
\usepackage{array,blindtext}

\newcommand\beq{\begin{equation}}
\newcommand\eeq{\end{equation}}
\newcommand\bea{\begin{eqnarray}}
\newcommand\eea{\end{eqnarray}}

\begin{document}
\title{Low Energy Spectrum of  SU(2) Lattice Gauge Theory: An Alternate Proposal via Loop Formulation}
\author{
Indrakshi Raychowdhury\footnote{indrakshi.raychowdhury@ulb.ac.be}\\
Center for Nonlinear Phenomena and Complex Systems,\\
Universit´e Libre de Bruxelles,\\
CP 231, Campus Plaine,
B-1050 Brussels,
Belgium}
\maketitle
\begin{abstract}
We show that,  prepotential formulation of gauge theories on honeycomb lattice yields local loop states, which are free from any spurious loop degrees of freedom and hence exact and orthonormal. We also illustrate that, the dynamics of orthonormal loop states are exactly same in both the square and honeycomb lattices. We further extend this construction to arbitrary dimensions. Utilizing this result, we make a mean field ansatz for loop configurations for SU(2) lattice gauge theory in $2+1$ dimension contributing to the low energy sector of the spectrum. Using variational analysis, we show that,  this type of mean loop configurations  has two distinct phases in the strong and weak coupling regime and shows a first order transition at $g=1$.  We then propose a reduced Hamiltonian to describe the dynamics of the theory within the mean field ansatz.  We further work with the mean loop configuration obtained at the weak coupling regime and analytically calculate the spectrum of the reduced Hamiltonian. The spectrum  matches with that of the existing literature in this regime, establishing our ansatz to be a valid alternate one which is far more easier to handle for computation.
\end{abstract}

\section{Introduction}

Understanding the low energy behaviour of gauge theories is one of the most important problem of particle physics. Formulation of gauge theories on discrete lattice \cite{wil} has shown many major investigations in this direction over past few decades. Using Monte Carlo method, many important physical quantities can be computed numerically \cite{Creutz}. However, understanding  the vacuum as well as excited states in this sector is still open for investigation. Hence, there should always be attempts  to make analytic approximations, specifically in the weak coupling regime of lattice gauge theory. This present work proposes such an approximation for SU(2) lattice gauge theories in $2+1$ dimensions, which can as well be generalized to higher dimensions and higher gauge groups. 

We work within the Hamiltonian formulation of lattice gauge theory \cite{kogut} and use prepotential \cite{pp} framework, which gives an useful reformulation in terms of gauge invariant loops. Loop formulation of gauge theories is always desired by theorists \cite{loops} as one can get rid of spurious gauge degrees of freedom. However, working with loop does not guarantee to work with only physical degrees of freedom as the loop space itself is highly over-complete \cite{mans}.
Working in terms of gauge invariant loops in the weak coupling regime is again particularly difficult, as all possible loops of all shapes and sizes do contribute to the low energy spectrum of weakly coupled gauge field theories. In this scenario, the prepotential formulation \cite{pp} gives a great advantage over the standard Wilson loop approach as it is possible to extract only physically relevant loop degrees of freedom and study the dynamics of those.

In prepotenntial formulation of lattice gauge theory \cite{pp}, one constructs gauge invariant loop variables, locally at each site. This particular feature makes it extremely useful for different directions of exploring the subject starting from analytic calculations \cite{prd,sreeraj,new} to the recent approach of quantum simulating lattice gauge theory \cite{qs}. However, the original motivation for formulating prepotential approach was to make better understanding at the weak coupling regime of the theory. Towards this direction, a very important step is to construct the exact and orthonormal loop Hilbert space, containing only physical degrees of freedom. In this work, (also in another parallel recent work by Anishetty et al. \cite{new}), we have proposed a general technique of constructing explicit orthonormal loop states for SU(2) lattice gauge theory in any arbitrary dimension in an extremely easy way even without going into the complicated Clebsch-Gordon coeffiecients. 
The prepotential formulation on square lattice reveals that the physically relevant loops are the non-intersecting ones \cite{prd}. We  propose that, if one virtually splits each site of the lattice into two virtual ones following figure \ref{ptsplit}, the resulting lattice is an hexagonal one in two dimension. It has also been demonstrated that the dynamics of orthogonal loop states on square lattice is exactly equivalent to that of all possible loops on hexagonal one. The prepotential formulation on hexagonal lattice  keeps all the important features of this particular formulation intact, such as local loop description by constructing intertwiners  and Abelian weaving of those intertwiners leads to standard Wilson loops of the theory \cite{pp}. Moreover, the extra advantage that is obtained on hexagonal lattice, is that the local loop space constructed at each site is exact and orthonormal. This is a tremendous advantage for the purpose of any computation be it analytic or numerical as one needs to work within   a really small Hilbert space without bothering about complicated Mandelstam constraint. In \cite{new}, it has also been shown that  there exists, more than one way to split each point, and the resulting lattices turn out to be of different types (i.e hexagon,octagon, square etc) as well as of different translational symmetries. It has also been argued in \cite{new} that, dynamics on these virtual lattices are exactly equivalent to that of the original lattice by gauge fixing on the virtual links connecting splitted lattice sites. One can choose any splitting scheme as per the calculational convenience.  For the purpose of present work we fix the splitting scheme given in figure \ref{ptsplit} at each site, and get a hexagonal lattice to work with. We establish its equivalence with the original lattice by computing the dynamics explicitly for random loop configurations generated on both of these.



In this work, we start from prepotential formulation of pure SU(2) gauge theory on square lattice. Then using the virtual point splitting technique of \cite{new}, we move to the virtual hexagonal lattice. On this particular lattice, we make a mean field ansatz  that only an average loop configuration contribute to the low energy spectrum of SU(2) lattice gauge theory. We also show that this average loop configuration has two distinct phases at the weak and strong coupling regime of the theory and shows a first order phase transition at $g=1$.
 We also construct a reduced Hamiltonian, starting from the Kogut-Susskind Hamiltonian, which keeps the dynamics of the theory within the mean field ansatz. We further perform a variational calculation, to fix the mean field configuration at different values of coupling. This analysis also shows that the average value of the fluxes flowing across each site shows a distinct jump as one moves from strong to weak coupling regime at $g=1$. We are however interested in the the loop configuration at small values of  $g$ . This analysis reveals that large fluxes contribute to the low energy spectrum at weak coupling regime as opposed to the zero flux at strong coupling regime. For the original Kogut-Susskind Hamiltonian, largest contribution should come from large loops carrying large fluxes. Working with prepotentials makes us free from considering large loops at all, as all loops are now local \cite{pp,prd,new}. Hence, our ansatz of the weak coupling vacuum consists of only large fluxes flowing across the sites. 
 We finally compute the lower lying spectrum of that reduced Hamiltonian at weak coupling regime and show that we get reasonably acceptable results within this approximation. 

The organization of the paper is as follows: in section 2 we briefly discuss the loop formulation of SU(2) gauge theory on hexagonal lattice and compare its dynamics with that of the square lattice. In section 3, we illustrate the  the origin of hexagonal lattice from square lattice by virtual point splitting in arbitrary dimension. In section 4, we discuss the average loop configuration for prepotential formulation of SU(2) theory in $2+1$ dimentions on virtual hexagonal lattice. In section 5, we propose  a reduced Hamiltonian and discuss its dynamics  within the mean field ansatz and finally compute low energy spectrum within this ansatz. Finally we summarize and discuss future aspects of this study in section 6.

\section{Loop formulation of SU(2) gauge theory on hexagonal lattice}

\begin{figure}[h]
\begin{center}
\begin{overpic}[scale=0.3,unit=5mm]{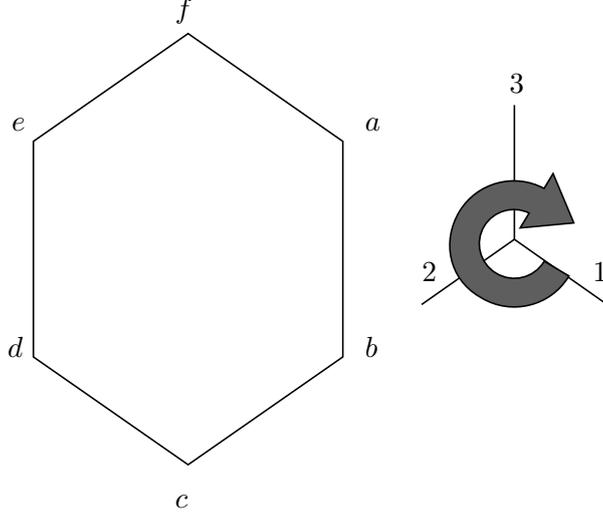}
\put(9.5,4){$b$}
\put(9.5,10){$a$}
\put(0.2,10){$e$}
\put(0.1,4){$d$}
\put(4.5,0){$c$}
\put(4.5,13){$f$}
\put(15.5,6){$1$}
\put(11,6){$2$}
\put(13.3,11){$3$}
\end{overpic}
\caption{A plaquette (surrounded by six vertices $a,b,c,d,e,f$) on hexagonal lattice. From each vertex, links emerge in directions $1,2,3$. The orientation denoted in this figure defines the convention to construct the local loop operators and states in (\ref{genloopst})} 
\label{hplaq}
\end{center}
\end{figure}
\begin{figure}[h]
\begin{center}
\includegraphics[width=0.5\textwidth,height=0.04\textwidth]
{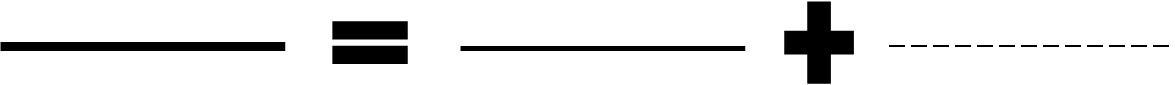}
\end{center}
\caption{Link operators consisting of two parts in prepotential, one of which increases flux along that link by one unit and another one decreases.} 
\label{link+-}
\end{figure}

Let us consider SU(2) pure gauge theory, formulated on 2 dimensional spatial lattice consisting of hexagonal plaquettes as shown in figure \ref{hplaq}.
As in the Kogut-Susskind Hamiltonian formulation \cite{kogut}, each of the links $(n,i)$ originating from site $n$ along direction $i$ \footnote{On hexagonal lattice, i=1,2,3 as denoted in figure \ref{hplaq}, although the physical dimension of lattice is only two.} carries a link variables $U(n,i)$ and there are SU(2) generators $E^{a}(n,i),~ a=1,2,3$ present at each end of the links. The Hamiltonian, in terms of these canonically conjugate variables reads as:
\bea
\label{ham_hex}
H_{KS}= g^2 \sum_{\mbox{links}} E^{2}_{links} + \frac{1}{g^2} \sum_{plaquettes}\left( 4- \mbox{Tr }U_{plaquette}- \mbox{Tr }U^\dagger_{plaquette}  \right)
\eea 
Within prepotential framework \cite{pp}, we attach e set of prepotential doublet (as we are working with SU(2)) $a^\dagger_{\alpha}(L) \& a^\dagger_{\alpha}(R)$  at left and right end of each link $(n,i)$ with $\alpha=1,2$ and $i=1,2,3$. In terms of preotentials \cite{pp}, the electric field is given by,
\bea
E_{L/R}^a= a^\dagger_{\alpha}(L/R)\left(\frac{\sigma^a}{2}\right)^\alpha_{\beta}a^\beta(L/R)
\eea
satisfying the SU(2) algebra at each end.
 The link variable at site $n$ along the direction $i$ takes the form given by:
\bea
\label{linkop}
U^{\alpha}{}_{\beta}(n,i)=\frac{1}{\sqrt{\hat N_i+1}}\left(\tilde{a}^{\dagger\alpha}(L) \, a^{\dagger}_\beta(R) 
+ a^\alpha(L)  \, \tilde{a}_\beta(R)\right)\frac{1}{\sqrt{\hat N_i+1}}
\eea
where, $N_i=a^\dagger(L)\cdot a(L)=a^\dagger(R)\cdot a(R)$ counts the number of prepotentials along direction $i$. 
From (\ref{linkop}), we find that, the link operator is a sum of a creation part, which increases the flux by one unit and another annihilation part, where flux is  decreased by same. Hence, pictorially we can think of the link as shown in figure \ref{link+-}.
 For hexagonal plaquette this decomposition yields $2^6$ plaquette terms (as opposed to $2^4$ terms in the case of square plaquette). 
Likewise on square lattice, the prepotential formulation on hexagonal lattice, yields  a local loop description of the theory which we will show to be free from any loop redundancy. Each of the plaquette operators basically consists of six local loop operators glued together following Abelian Gauss law.
 Let us first concentrate on the local loop operators and loop states constructed on at each site of the hexagonal plaquette.
 
\begin{figure}[h]
\begin{center}
\includegraphics[width=0.5\textwidth,height=0.2\textwidth]
{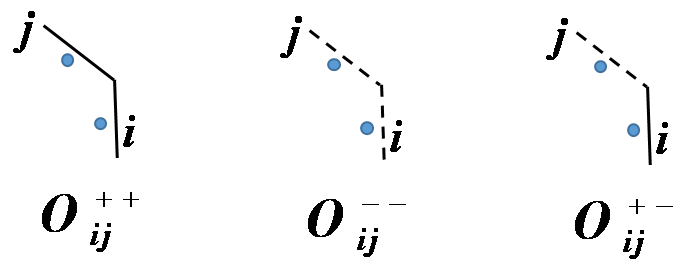}
\end{center}
\caption{pictorial representation of all possible local loop operators at each site of hexagonal lattice. A solid dot on solid line denotes the creation operator along that link and on the dashed line denotes annihilation operator. For hexagonal lattice $i,j=1,2,3$.} 
\label{hlocalloopop}
\end{figure}

\begin{figure}[h]
\begin{center}
\begin{overpic}[scale=0.2,unit=5mm]{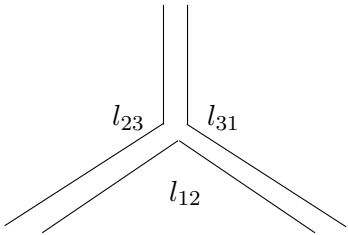}
\put(5.5,2){$l_{12}$}
\put(6.5,4){$l_{31}$}
\put(4,4){$l_{23}$}
\end{overpic}
\caption{A general loop state at a site of hexagonal lattice.} 
\label{localloop}
\end{center}
\end{figure}

At each site of the hexagonal lattice, links emerge in three directions, labelled as $1,2,3$. Each link is associated with a prepotential creation and annihilation operators $\{a^\alpha(n,i), a^\dagger_{\alpha}(n,i)\}$ for $i=1,2,3$ and $\alpha=1,2$. The gauge invariant operator constructed out of these are:
\bea
\mathcal O^{++}_{ij} &\equiv & \epsilon^{\alpha\beta}a^\dagger_{\alpha}(n,i)a^\dagger_{\beta}(n,j) \label{o++} \\
\mathcal O^{--}_{ij} &\equiv & \epsilon_{\alpha\beta}a^{\alpha}(n,i)a^{\beta}(n,j) \label{o--} \\
\mathcal O^{+-}_{ij} &\equiv & a^\dagger_{\alpha}(n,i)a^{\alpha}(n,j) \label{o+-}
\eea
where, $i,j$ are the direction indices with $i\ne j$. These local loop operators on the hexagonal lattice are represented pictorially in figure \ref{hlocalloopop}.
 It is clear from the above set of equations that, acting on strong coupling vacuum only the first operator, i.e the one given in (\ref{o++}) will give non-zero contribution and will build up the local loop Hilbert space. For two dimensional hexagonal lattice, we characterize the local loop space by three independent linking numbers $l_{ij}$ denoting the flux flowing along three $ (ij)$ directions, namely $(12), (23)$ and $(31)$ as shown in figure \ref{localloop}.

  Hence, the most general loop state at each site is given by: 
\bea
|l_{12},l_{23},l_{31}\rangle_x= \mathcal N \prod_{i\ne j|_x} \left( \mathcal O^{++}_{ij}\right)^{l_{ij}}|0\rangle_x
\label{genloopst}
\eea
where, $\mathcal N$ is the normalization factor.
Note that, on hexagonal lattice, only three linking numbers are present at each  characterizing a complete basis. The local and orthonormal loop states at each site is explicitly obtained as,
\bea
\label{lij}
|l_{12},l_{23},l_{31}\rangle = \frac{1}{(l_{12}+l_{23}+l_{31})! ( l_{12} )! ( l_{23} )! ( l_{31} )!} \left( \mathcal O^{++}_{12}\right)^{l_{12}}\left( \mathcal O^{++}_{23}\right)^{l_{23}}\left( \mathcal O^{++}_{31}\right)^{l_{31}}|0\rangle
\eea
The $|l_{ij}\rangle$ basis on hexagonal lattice is exactly equivalent to the number operator basis, where three number operators $n_1,n_2,n_3$ counts the number of prepotentials along each direction. These two basis are related as:
\bea
&& \label{l2n}\hspace{-2cm} n_1=l_{12}+l_{31} ~~, ~~ n_2=l_{12}+l_{23} ~~,~~ n_3=l_{23}+l_{31} \\
\mbox{or equivalently,} && l_{12}=\frac{1}{2}(n_1+n_2-n_3) ~,~ \nonumber \\ && l_{23}=\frac{1}{2}(n_2+n_3-n_1) ~,~\label{n2l} \\ && l_{31}=\frac{1}{2}(n_1+n_3-n_2)\nonumber
\eea
From the above set of equations, note that, the $l_{ij}$ variables are truly independent positive integers ranging from zero to infinity, whereas the transformed $n_i$ variables not truly independent but derived from $l_{ij}$ basis following (\ref{l2n}). Unlike $l_{ij}$ variables, the ranges of $n_i$'s are restricted to satisfy the triangle inequalities given in (\ref{n2l}) keeping the $l_{ij}$'s positive semi-definite. In this regard, working with the $l_{ij}$ basis is more convenient as it involve no further constraint unlike the $n_{i}$ basis, however one can always move from one to another following (\ref{l2n}) or (\ref{n2l}).
\begin{figure}[h]
\begin{center}
\includegraphics[width=0.9\textwidth,height=0.3\textwidth]
{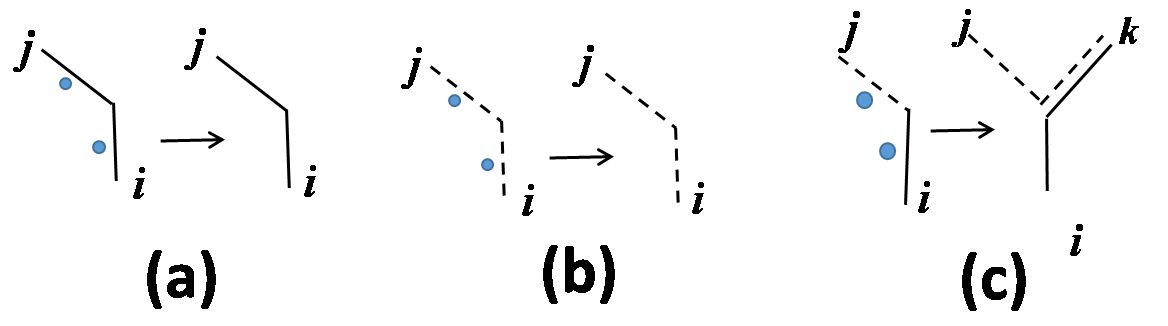}
\end{center}
\caption{pictorial representation of the action of local loop operators on local loop states of hexagonal lattice. In each of these actions, the left hand side denotes the operator (denoted by dot on solid or dashed line) acting on a state $|l_{12},l_{23},l_{31}\rangle$ and the right hand side denotes the resultant state produced where the particular fluxes along a particular direction is either increased (denoted by solid line) or decreased (denoted by dshed line). } 
\label{loopac}
\end{figure}
Next is to find out the action of the local loop operators (\ref{o++},\ref{o--},\ref{o+-})  on orthonormal local loop states (\ref{lij}). These actions are far more simple than that in case of square lattice,  derived in \cite{prd}. The first and simplest operator (\ref{o++}) involves no annihilation operator, and hence simply gives:
\bea
\label{ac++}
\mathcal O^{++}_{ij} |l_{ij}\rangle = \sqrt{(l_{ij}+1)(l_{12}+l_{23}+l_{31}+2)}|l_{ij}+1\rangle
\eea
Note that, the coefficient is obtained to create another normalized state from the one on which it acts. This loop action is pictorially represented in figure \ref{loopac}(a) showing that the resultant flux along $ij$ direction increases by one unit. 

The next loop operator in (\ref{o--}) involves only annihilation operator and its action on a general loop state is:
\bea
\label{ac--}
\mathcal O^{--}_{ij} |l_{ij}\rangle = \sqrt{l_{ij}(l_{12}+l_{23}+l_{31}+1)}|l_{ij}-1\rangle
\eea
The right hand side of (\ref{ac--}) is obtained by shifting both the annihilation operators of $\mathcal O^{--}_{ij}$ to the right using commutation relations and adjusting the normalization factor as given in (\ref{lij}).

Another loop operator (\ref{o+-}) involves one creation as well as one annihilation operator and in the very same way its action on arbitrary loop state is obtained as:
\bea
\label{ac+-}
\mathcal O^{+-}_{ij} |l_{ij}\rangle = -\sqrt{(l_{ik}+1)l_{jk}}|l_{jk}-1,l_{ik}+1\rangle
\eea
Note that, as there are only three links emerging from each site, for a given operator $\mathcal O^{+-}_{ij}$, the direction $k$, on which there will be change in flux configuration, but no change in net flux is always fixed, and hence there is no summation over index $k$, in the right hand side of (\ref{ac+-}). 
The last two loop actions (\ref{ac--},\ref{ac+-}) are also represented pictorially in figure \ref{loopac} (b) and (C) respectively.

\begin{figure}[h]
\begin{center}
\begin{overpic}[scale=0.6,unit=1mm]{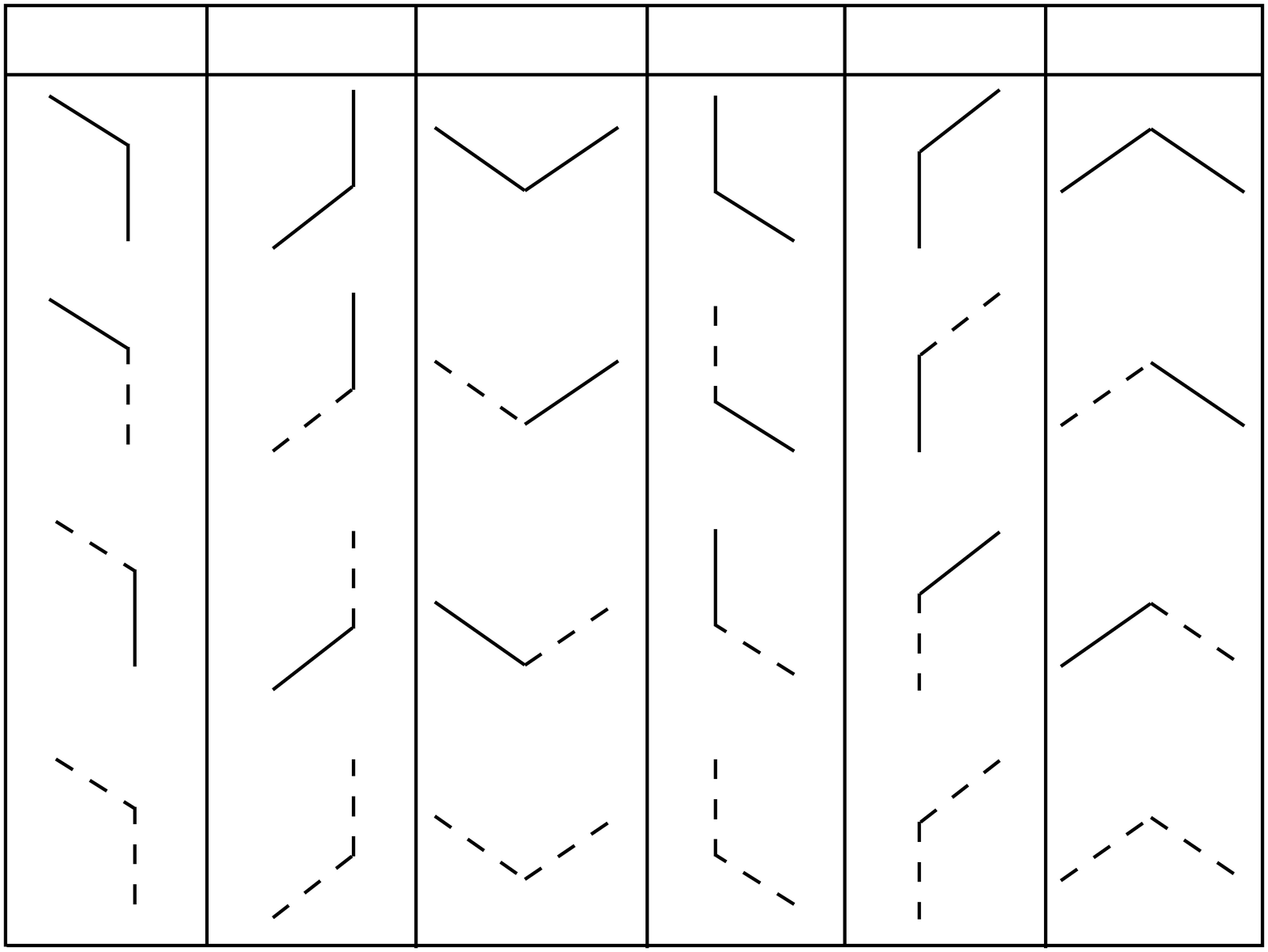}
\put(9,120){At vertex `$a$'}
\put(33,120){At vertex `$b$'}
\put(60,120){At vertex `$c$'}
\put(90,120){At vertex `$d$'}
\put(116,120){At vertex `$e$'}
\put(140,120){At vertex `$f$'}
\put(8,105){$\mathcal O^{++}_{31}$}
\put(8,79){$\mathcal O^{-+}_{31}$}
\put(8,52){$\mathcal O^{+-}_{31}$}
\put(8,19){$\mathcal O^{--}_{31}$}
\put(34,105){$\mathcal O^{++}_{23}$}
\put(34,79){$\mathcal O^{-+}_{23}$}
\put(34,52){$\mathcal O^{+-}_{23}$}
\put(34,19){$\mathcal O^{--}_{23}$}
\put(68,110){$\mathcal O^{++}_{12}$}
\put(68,79){$\mathcal O^{-+}_{12}$}
\put(68,50){$\mathcal O^{+-}_{12}$}
\put(68,22){$\mathcal O^{--}_{12}$}
\put(96,105){$\mathcal O^{++}_{31}$}
\put(96,79){$\mathcal O^{-+}_{31}$}
\put(96,52){$\mathcal O^{+-}_{31}$}
\put(96,19){$\mathcal O^{--}_{31}$}
\put(123,105){$\mathcal O^{++}_{23}$}
\put(123,77){$\mathcal O^{-+}_{23}$}
\put(123,48){$\mathcal O^{+-}_{23}$}
\put(121,18){$\mathcal O^{--}_{23}$}
\put(146,104){$\mathcal O^{++}_{12}$}
\put(146,73){$\mathcal O^{-+}_{12}$}
\put(146,43){$\mathcal O^{+-}_{12}$}
\put(146,16){$\mathcal O^{--}_{12}$}
\end{overpic}\\
\label{hexloopop}
\caption{Pictorial representation of all possible loop actions around a plaquette `$abcdef$' given in figure \ref{hplaq}}
\end{center}
 \end{figure}

Let us now concentrate to the complete plaquette, around which the Hamiltonian dynamics evolves. As stated earlier, plaquettes are the smallest Wilson loops and the basic gauge invariant variables to appear in the Kogut-Susskind Hamiltonian. But in terms of prepotentials, it is not the fundamental one but consists of six vertices as shown in figure \ref{hplaq} each consisting of the full set of loop operators and states. All possible loop operators at each vertex which are given in figure \ref{hexloopop}, which are of type listed in (\ref{o++}-\ref{o+-}). Their action on loop state as derived above is listed in detail in table 1 for convenience. 

\begin{table}
\label{tablecoeff}
\begin{center}
\begin{tabular}{|c|c|c|c|}
\hline
Sl no. & At vertices & \mbox Action on $|l_{12},l_{23},l_{31}\rangle$ & Coefficient \\  
\hline 
1. & $c$ and $f$ & $\mathcal O^{++}_{12}|l_{12},l_{23},l_{31}\rangle =C_1|l_{12}+1\rangle$ & $C_1=\sqrt{(\frac{l_{12}+1)(l_{12}+l_{23}+l_{31}+1)}{(l_{12}+l_{31}+2)(l_{12}+l_{23}+1)}}$ \\
\hline 2.  & $c$ and $f$ &  $\mathcal O^{-+}_{12}|l_{12},l_{23},l_{31}\rangle =-C_2|l_{23}+1,l_{31}-1\rangle$ & $C_2=\sqrt{\frac{l_{31}(l_{23}+1)}{(l_{12}+l_{31}+2)(l_{12}+l_{23}+1)}}$ \\
\hline 3. & $c$ and $f$ & $\mathcal O^{+-}_{12}|l_{12},l_{23},l_{31}\rangle =-C_3|l_{23}-1,l_{31}+1\rangle$ & $C_3=\sqrt{\frac{l_{23}(l_{31}+1)}{(l_{12}+l_{31}+1)(l_{12}+l_{23}+2)}}$ \\
\hline 4.& $c$ and $f$ & $\mathcal O^{--}_{12}|l_{12},l_{23},l_{31}\rangle =C_4|l_{12}-1\rangle$ & $C_4=\sqrt{\frac{l_{12}(l_{12}+l_{23}+l_{31}+1)}{(l_{12}+l_{31}+1)(l_{12}+l_{23}+2)}}$  \\
\hline 5. & $a$ and $d$ & $\mathcal O^{++}_{31}|l_{12},l_{23},l_{31}\rangle =C_5|l_{31}+1\rangle$ &$C_5=\sqrt{\frac{(l_{31}+1)(l_{12}+l_{23}+l_{31}+2)}{(l_{23}+l_{31}+1)(l_{12}+l_{31}+2)}}$ \\
\hline 6. & $a$ and $d$ & $\mathcal O^{-+}_{31}|l_{12},l_{23},l_{31}\rangle =-C_6|l_{12}+1,l_{23}-1\rangle$ & $C_6=\sqrt{\frac{l_{23}(l_{12}+1)}{(l_{23}+l_{31}+2)(l_{12}+l_{31}+1)}}$ \\
\hline 7.  & $a$ and $d$  & $\mathcal O^{+-}_{31}|l_{12},l_{23},l_{31}\rangle =-C_7|l_{12}-1,l_{23}+1\rangle$ & $C_7=\sqrt{\frac{l_{12}(l_{23}+1)}{(l_{23}+l_{31}+1)(l_{12}+l_{31}+2)}}$ \\
\hline 8. & $a$ and $d$ & $\mathcal O^{--}_{31}|l_{12},l_{23},l_{31}\rangle =C_8|l_{31}-1\rangle$ & $C_8=\sqrt{\frac{l_{31}(l_{12}+l_{23}+l_{31}+1)}{(l_{23}+l_{31}+1)(l_{12}+l_{31}+2)}}$  \\
\hline 9.  & $b$ and $e$ & $\mathcal O^{++}_{23}|l_{12},l_{23},l_{31}\rangle =C_9|l_{23}+1\rangle$ & $C_9=\sqrt{\frac{(l_{23}+1)(l_{12}+l_{23}+l_{31}+2)}{(l_{23}+l_{12}+1)(l_{23}+l_{31}+2)}}$  \\
\hline 10.  & $b$ and $e$  & $\mathcal O^{-+}_{23}|l_{12},l_{23},l_{31}\rangle =-C_{10}|l_{31}+1,l_{12}-1\rangle$ & $C_{10}=\sqrt{\frac{l_{31}(l_{12}+1)}{(l_{23}+l_{31}+2)(l_{23}+l_{12}+1)}}$ \\
\hline 11.  & $b$ and $e$  & $\mathcal O^{+-}_{23}|l_{12},l_{23},l_{31}\rangle =-C_{11}|l_{31}-1,l_{12}+1\rangle$ & $C_{11}=\sqrt{\frac{l_{12}(l_{31}+1)}{(l_{23}+l_{31}+1)(l_{23}+l_{12}+2)}}$ \\
\hline 12. & $b$ and $e$  & $\mathcal O^{--}_{23}|l_{12},l_{23},l_{31}\rangle =C_{12}|l_{23}-1\rangle$ & $C_{12}=\sqrt{\frac{l_{23}(l_{12}+l_{23}+l_{31}+1)}{(l_{23}+l_{12}+1)(l_{23}+l_{31}+2)}}$ \\
\hline
\end{tabular}
\caption{ Loop actions around the hexagonal plaquette, given in figure \ref{hexloopop}. The coefficients in the loop action  are given in the last column.}
\end{center}
 \end{table}

One important thing to note here is that, as per our convention of defining loop operators and loop states, some of the loop  actions bring $-$ve sign in the coefficient as shown in table \ref{tablecoeff}. But this happens only for the mixed operators (i.e type $\mathcal O^{+-}$) which involves one creation and one annihilation operator. But when we consider a closed loop such as a plaquette, we see that these type of mixed vertices always appear in pairs. Moreover note that, for the full plaquette operator (or any closed loop), the mixed terms can only appear in pairs and hence, the plauette operators (or any closed loop operators) are always positive by our convention. 

Further note that, the action of loop operators on any loop state consists of two parts, one contains a numerical coefficient or number operators and another is some shift operators for the linking numbers.  The coefficient  that appears in table 1 are calculated follows from the convention that, the shift operators are always at right most position and coefficients (function of number operators) are at left. 
 
Having set the action of the loop operators on arbitrary loop states, one can easily compute the matrix element of the magnetic part of the Hamiltonian (\ref{ham_hex}) within orthonormal loop basis, characterized by $l_{12}, l_{23},l_{31}(\mbox{or, } n_1,n_2,n_3)$ basis. The magnetic Hamiltonian consists of $2^6=64$ terms, each of which is a set of six local loop operator at each of the six vertices of the hexagon, the action of which on respective local loop states are computed following the table \ref{tablecoeff}.

\subsection{Dynamics on hexagonal lattice vs. dynamics on square lattice}

 At this point we compare the Hamiltonian dynamics on hexagonal lattice with that on the square lattice  numerically. For this purpose, we generate an arbitrary but valid loop configuration around one particular hexagonal plaquette of the lattice. This is done by specifying a set of three positive semi-definite integers denoting $n_{1},n_{2},n_{3}$ at each of the alternate site (say at site a, c and e of the plaquette in figure \ref{hplaq}). We denote these alternate sites as the even sites of the lattice. The set of three integers when satisfy triangle inequalities 
\bea
n_i+n_j\ge n_k~~~~~~ \forall i\ne j\ne k
\eea
are accepted as a valid loop configuration. 
For the neighbouring odd sites, i.e for sites b,d and f,  two numbers are fixed by even sites a, c, e,  and the third one is generated randomly satisfying triangle inequalities. The prefixed ones are:
\bea
n_1(b)=n_1(a) ~&,& ~ n_3(b)=n_3(c)\nonumber \\
n_2(d)=n_2(c) ~&,& ~ n_1(d)=n_1(e)\nonumber \\
n_3(f)=n_3(e) ~&,& ~ n_2(f)=n_2(a)
\eea
and randomly generate $n_2(b), n_3(d) ~\& ~ n_1(f)$ satisfying triangle inequalities at b,d and f sites as well. \footnote{ Fixing the loop quantum numbers $n_1,n_2,n_3$ at all the even sites throughout the lattice, automatically  fixes the configurations at all of the odd sites. A valid loop configuration is obtained if triangle inequality is valid at each and every sites. However, for that case generating linking numbers throughout the lattice and picking valid configurations when Abelian Gauss law (\ref{u1gl}) is satisfied is another option. It requires a detail study to find out which one is the most efficient one. Here, as we are only interested in dynamics around a chosen plaquette, we do not bother to generate loop configurations throughout the lattice.} Having generated a loop configuration denoted by $n_1,n_2,n_3$ at each of the six sites around the plaquette, we readily compute the corresponding $l_{12},l_{23}, l_{31}$ at each of them following (\ref{n2l}).
All possible loop operators residing at each of the six vertices around the hexagonal plaquette, changes the loop configurations following a coefficient listed in table \ref{tablecoeff}.  The dynamics of an arbitrary hexagonal plaquette in compact form is given below:
\bea
\label{hexdyn}
&& \langle \bar j_1^h, \bar j_2^h, \bar j_3^h, \bar j_4^h, \bar j_5^h, \bar j_6^h| \left(\mbox{Tr} U_{plaquette}+ \mbox{Tr} U^\dagger_{plaquette}\right)| j_1^h, j_2^h, j_3^h, j_4^h, j_5^h, j_6^h \rangle\nonumber \\
&=&  \left(\mathcal C_1^+\delta_{\bar j_1^h,j_1^h+\frac{1}{2}} + \mathcal C_1^-\delta_{\bar j_1^h,j_1^h-\frac{1}{2}} \right)\left(\mathcal C_2^+\delta_{\bar j_2^h,j_2^h+\frac{1}{2}} + \mathcal C_2^-\delta_{\bar j_2^h,j_2^h-\frac{1}{2}} \right)\nonumber \\
&& \left(\mathcal C_3^+\delta_{\bar j_3^h,j_3^h+\frac{1}{2}} + \mathcal C_3^-\delta_{\bar j_3^h,j_3^h-\frac{1}{2}} \right) \left(\mathcal C_4^+\delta_{\bar j_4^h,j_4^h+\frac{1}{2}} + \mathcal C_4^-\delta_{\bar j_4^h,j_4^h-\frac{1}{2}} \right)\nonumber \\
&& \left(\mathcal C_5^+\delta_{\bar j_5^h,j_5^h+\frac{1}{2}} + \mathcal C_5^-\delta_{\bar j_5^h,j_5^h-\frac{1}{2}} \right)\left(\mathcal C_6^+\delta_{\bar j_6^h,j_6^h+\frac{1}{2}} + \mathcal C_6^-\delta_{\bar j_6^h,j_6^h-\frac{1}{2}} \right)
\eea 
where, $\mathcal C_i^{\pm}$ for $i=1,2,...,6$ 's are some algebraic coefficients which are functions of number operators. In the full plaquette operator, they always come in pairs as in the six vertices of the plaquette. Product of two such $\mathcal C_i^{\pm}$s are the vertex coefficient $C_j$ for $j=1,2,..,12$ listed in table \ref{tablecoeff}. We compute those coefficeints for our configuration to find out the matrix element $(M_h)$ of magnetic Hamiltonian between a  randomly selected initial and final states,  few of which are  listed in table \ref{sqhexcomptab}. 


Having calculated numerical values of the matrix elements of the magnetic Hamiltonian for hexagonal plaquette, we will now compare those matrix elements with that for a square plaquette (Matrix elements of magnetic Hamiltonian on a square lattice is given in Appendix A). 

\begin{figure}[h]
\begin{center}
\begin{overpic}[scale=0.3,unit=5mm]{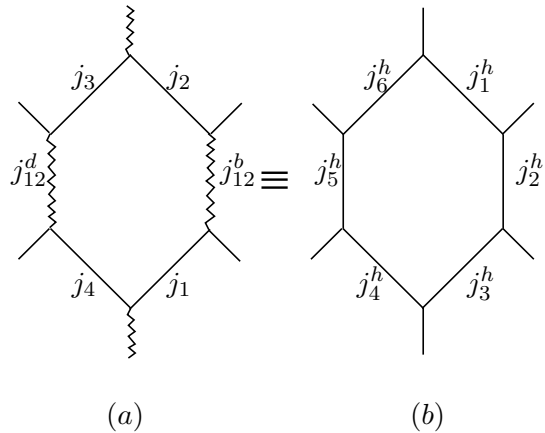}
\put(12.5,10){$j_1^h$}
\put(9.8,10){$j_6^h$}
\put(4.6,10){$j_2$}
\put(2.1,10){$j_3$}
\put(12.5,4.5){$j_3^h$}
\put(9.6,4.5){$j_4^h$}
\put(4.6,4.5){$j_1$}
\put(2.1,4.5){$j_4$}
\put(13.8,7.5){$j_2^h$}
\put(8.5,7.5){$j_5^h$}
\put(6.0,7.5){$j_{12}^b$}
\put(0.5,7.5){$j_{12}^d$}
\put(3,1){$(a)$}
\put(11,1){$(b)$}
\end{overpic}
\end{center}
\caption{(a) orthonormal fluxes, that change around a square plaquette as given in (\ref{dyna1}); (b) fluxes around an hexagonal plaquette. Both in (a) and (b), the fluxes on the external links has not been written explicitly as they remain same in the particular plaquette interaction, as evident in (\ref{dyna1}) and (\ref{hexdyn}) respectively.} 
\label{squarehexcomp}
\end{figure}

An important point to note at this point is,  even for a square plaquette, the loop configuration actually changes over a virtual hexagonal plaquette consisting of four real fluxes flowing around a plaquette together with two internal fluxes (coupled angular momenta, see Appendix A for detail).
We  provide the dictionary of identifying square plaquette to hexagonal plaquette in figure \ref{squarehexcomp}. 

The dictionary for shifting between square (given in appendix A) and hexagonal plaquette (given in section 2) is as follows:
\begin{itemize}
\item Each site $(s\equiv a,b,c,d)$ on the square lattice has the following orthonormal angular momentum to have net angular momentum zero at site $x$:
$$j_1^x,j_2^x,j_{\bar 1}^x,j_{\bar 2}^x,j_{12}^x=j_{\bar 1\bar 2}^s$$
\item Identify the flux around a square plaquette $abcd$ as:
\bea
j_1^a=j_{\bar 1}^b\equiv j_1 ~& ~j_2^b=j_{\bar 2}^c\equiv j_2 \nonumber \\
j_{\bar 1}^c=j_{ 1}^d\equiv j_{\bar 1}  ~&~ j_{\bar 2}^d=j_{2}^a\equiv j_{\bar 2} \nonumber
\eea 
\item At each site $(s\equiv a,b,c,d,e,f)$ of the hexagonal plaquette, the orthonormal states are characterized by $l_{12}^x,l_{23}^x,l_{31}^x$ or equivalently by $$n_1^x\equiv 2j_1^h|_x, ~n_2^x\equiv 2j_2^h|_x, ~n_3^x\equiv 2j_3^h|_x$$ following (\ref{l2n})\footnote{Note that, the hexagonal lattice contains alternate odd (b,d,f) and even (a,c,e) sites. The links emerge  emerge in the direction $1,2,3$ from even sites and  in $\bar 1, \bar 2,\bar 3$ from odd sites. However, for most of the purposes, we will not differentiate even and odd sites in general and will consider links to emerge from all sites in direction $1,2,3$}.
\item Identify the flux (marked with `h') around the hexagonal plaquette `abcdef' as:
\bea
&& j_1^h=\frac{n_1^a}{2}=\frac{n_1^f}{2} ~~,~~
 j_2^h=\frac{n_3^a}{2}=\frac{n_3^b}{2} ~~,~~
  j_3^h=\frac{n_2^b}{2}=\frac{n_2^c}{2} ~~,
 \nonumber \\ &&
  j_4^h=\frac{n_1^c}{2}=\frac{n_1^d}{2} ~~,~~
   j_5^h=\frac{n_3^d}{2}=\frac{n_3^e}{2}  ~~,~~
    j_6^h=\frac{n_2^e}{2}=\frac{n_2^f}{2} ~~.
\eea
\item Now, the same flux around the hexagonal plaquette, can be identified as the dynamic  flux around square plaquette `abcd'  in the following way:
\bea
j_1^h=j_2 ~,~ j_2^h=j_{12}^b~,~ j_3^h=j_1~,~ j_4^h=j_4~,~ j_5^h=j_{12}^d~,~ j_6^h=j_{3}
\eea
\item Moreover, the external links of plaquettes `abcd' and `abcdef' can also be identified, but we are not writing them explicitly as we find them to remain unchanged in this particular plaquette dynamics. 
\end{itemize}
Having established the connection between the particular square and hexagonal plaquette of interest we can now compute the matrix elements for magnetic Hamiltonian for both the cases. 

We now compare the above calculated dynamics around an hexagonal plaquette with that of the square plaquette given in (\ref{dyna1}). For this purpose we identify the fluxes around the hexagonal plaquette with those around a square plaquette as given in figure \ref{squarehexcomp}.
To compare the dynamics on square lattice and that on the hexagonal lattice, we simulate a random loop configuration on hexagonal lattice and compute the matrix element of the magnetic Hamiltonian as discussed above. Next we identify the same loop configuration on square lattice following prescription listed  above, and compute the matrix element of Magnetic Hamiltonian on square lattice for this state following (\ref{dyna1}).
 In this comparison it is easy to observe that the nontrivial delta functions in (\ref{hexdyn}) are exactly same as those arising in evaluating  the 6j symbols in (\ref{dyna1}). More importantly, our numerical calculation using random loop configuration reveals that the numerical value of the non-zero matrix elements for each and every cases matches exactly (upto a sign) with each other for the calculations done on square lattice and hexagonal lattice. The discrepancy in sign arises as for the particular convention of defining the loop states on square plaquette, that we have chosen, each and every term becomes positive. We repeat this comparison for 1000 random loop configurations and find this exact matching for each and every case. For the purpose of illustration, we only quote  a few sample results in table 2. 
\begin{table}
\begin{center}
\begin{tabular}{|c|c|c|c|}
\hline
 Initial state & Final state & Square plaquette & Hexagonal plaquette  \\
$|2j_1^h,2j_2^h,2j_3^h,2j_4^h,2j_5^h,2j_6^h\rangle $  &   $\{2\bar j_1^h,2\bar j_2^h,2\bar j_3^h,2\bar j_4^h,2\bar j_5^h,2\bar j_6^h\}$ & $M_s$ & $M_h$ \\
\hline   $|  8,15,13,9,5,11\rangle $  & $|  9,16,14,10,6,12\rangle $ & 3.682168E-002 & 3.682168E-002\\
\hline   $|  9,7,13,17,14,6\rangle $  & $| 10,6,14,18,15,7\rangle $ & 1.097742E-002 & 1.097742E-002\\
\hline   $|  9,16,12,14,14,7\rangle $  & $| 8,17,13,13,15,8\rangle $ & 1.350154E-002 & 1.350154E-002\\
\hline   $|  12,9,2,9,10,7\rangle $  & $| 11,8,3,10,9,8\rangle $ & 8.383834E-002 & 8.383834E-002\\
\hline   $|  7,10,6,10,10,16\rangle $  & $| 8,9,5,11,9,15\rangle $ & 4.790649E-002 & 4.790649E-002\\
\hline   $|  13,6,7,14,6,7\rangle $  & $| 12,5,6,13,5,8\rangle $ & 1.420527E-002 & 1.420527E-002\\
\hline
\end{tabular}
\caption{Few sample results for the comparison between dynamics around square plaquette and hexagonal plaquette. The first and second and column denotes two loop states around a hexagonal plaquette between which there exist a non zero matrix element for plaquette term $\mbox{Tr}\, U_{plaquette}$ of (\ref{ham_hex}). These configuration has been translated to a valid orthonormal loop configuration around a square plaquette using the prescription given in this section and the matrix element $M_s$ is calculated following Appendix A.  The same $M_h$ is also calculated for hexagonal plaquette following (\ref{hexdyn}).  Results are shown upto 8 decimal places upto which these two matches exactly. Note, these are only few sample results from numerical simulation.}
\label{sqhexcomptab}
\end{center}
 \end{table}
Hence, this numerical study proves that the dynamics of loop states on a square plaquette is identical to that on  an hexagonal plaquette as long as one is interested only in orthonormal loop states, which are actually relevant for exact physical degrees of freedom.

\section{Point Splitting and virtual hexagonal lattice}

In the last section, we have established the equivalence in dynamics of orthonormal loop configurations in a square and hexagonal lattice. In this section, we prescribe a virtual point splitting technique, which translates any square lattice to its hexagonal counter part. As a result of this transition, we gain a theory formulated in terms of only explicitly orthonormal loop degrees of freedom at each site, and pay the price of an extra Abelian Gauss law constraint. This price is actually negligible as the square lattice already had the Abelian constraints to solve and the extra one in hexagonal lattice is on very same footing as those. 

\begin{figure}[h]
\begin{center}
\begin{overpic}[scale=0.3,unit=5mm]{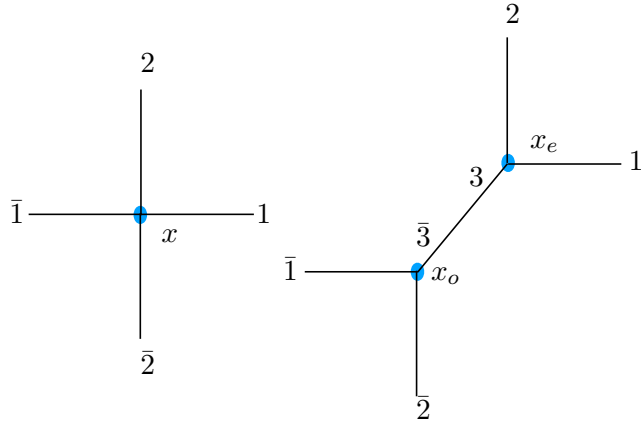}
\put(3.8,6){$x$}
\put(6.3,6.5){$1$}
\put(-0.2,6.5){$\bar 1$}
\put(3.25,10.5){$2$}
\put(3.25,2.5){$\bar 2$}
\put(10.9,5){$x_o$}
\put(13.5,8.5){$x_e$}
\put(16.1,7.9){$1$}
\put(7,5){$\bar 1$}
\put(12.85,11.8){$2$}
\put(10.5,1.3){$\bar 2$}
\put(11.9,7.5){$3$}
\put(10.5,5.9){$\bar 3$}
\end{overpic}
\end{center}
\caption{One site `$x$' on a square lattice is virtually splitted into two sites `$x_e~\& ~ x_o$' connected by a third virtual direction $3-\bar 3$ }
\label{ptsplit}
\end{figure}

The two dimensional  hexagonal lattice, that we demonstrated here, is  obtained by virtually splitting of each lattice site of a two dimensional square lattice. Elaborating a bit, consider a lattice site `x' as shown in figure \ref{ptsplit}. 
From this site on a 2d lattice, clearly 4 links emerge in  $1,2,\bar 1,\bar 2$ directions each carrying $n_i, ~(i=1,2,\bar 1,\bar 2)$ number of prepotentials (or angular momentum fluxes). We split this site `x' into two sites `$x_e ~\& ~x_o$'. Links from direction $1~\& ~2$ meet at site $x_e$, where as links from direction $\bar 1~\&~\bar 2$ meet at $x_o$. The splitted sites $x_1$-$x_2$ are connected by a virtual link along direction $3$. This same splitting done at each and every sites on the square lattice lifts the lattice to an hexagonal structure as shown in figure \ref{sq2h}. 
 
Now, prepotential formulation on this hexagonal lattice yields a local loop formulation of lattice gauge theory, exactly eqiuivalent to the  original square lattice, but contains only orthonormal and physical loop degrees of freedom as this is free from complicated Mandelstam constraints.  This makes the analysis on hexagonal lattice simpler for practical purpose of analytical as well as numerical computation.

At this point we must explicitly match the degrees of freedom of these two systems as well. For the site $x$ on the square lattice, there is six linking numbers or six loop degrees of freedom as earlier. Moreover, there was two Abelian Gauss law along two directions of the lattice (\ref{u1gl}) and one Mandelstam constraint (implying only non-intersecting loops contribute to physical degrees of freedom ). Resulting only three physical degrees of freedom. Now, coming back to hexagonal lattice, two sites, say $x_1 ~\&~x_2$ corresponds to actual site $x$ on the square lattice and  together should have only three degrees of freedom. Each site of the hexagonal lattice contains three linking numbers or three loop degrees of freedom, hence total six loop degrees of freedom matches with that of the square plaquette case. Unlike square plaquette, here there is no Mandelstam constraint at all to solve as there is no concept of intersecting loops. Together with the two Abelian Gauss law constraints along  directions $1$ and $2$ (same as square plaquette), there exists one more Abelian Gauss law constraint on the link connecting two splitted site (along direction $3$) and hence yielding exact degrees of freedom as square plaquette.  Hence, counting of degrees of freedom goes as:
\bea
\label{dof}
d.o.f=3 S-d-d_v
\eea
where, $S$ is the splitting index, which denotes each site has been splitted into $S$ virtual sites; $d$ is the dimension of the lattice, and $d_v$ is the number of virtual links connecting the splitted site, which is $d_v=S-1$. It is straightforward to show that this analysis smoothly extends to any higher dimension as well. i.e for $d$ dimensional spatial lattice, where $2d$ links meet at a site $x$, one can split the site into $S$ number of 3 point vertices. Obviously this splitting will results into $S-1$ number of intermediate links. However, the physical degrees of freedom for the site $x$ should still be $3(d-1)$. Equating both the sides of (\ref{dof}) yields $S=2(d-1)$.  Hence, this splitting of each site depends on the dimension we are working on. As we have already seen, for two dimension each lattice site splits into two virtual lattice sites, and for three dimensional case, it splits into 4 sites. However, in each case one can just work with three orthogonal loop state at each site and impose Abelian Gauss law on all of the links present. This simple analysis would result working with orthonormal loop basis for SU(2) lattice gauge theory in any arbitrary dimensions without getting involved with complicated Clebsch Gordon coefficients.

\section{The Hamiltonian and Average loop configurations}

In this work, we consider pure SU(2) gauge theory defined on a square lattice. The system is described by the Kogut-Susskind Hamiltonian,
\bea
\label{ham}
H_{KS}= g^2 \sum_{\mbox{links}} E^{2}_{links} + \frac{1}{g^2} \sum_{plaquettes}\left( 4- \mbox{Tr }U_{plaquette}- \mbox{Tr }U^\dagger_{plaquette}  \right)\equiv g^2H_E+\frac{1}{g^2}H_{mag}
\eea
The virtual point splitting, discussed in last section now enables us to define the  theory on hexagonal lattice. 

 On hexagonal lattice, the hexagonal plaquettes are surrounded by six links, out of which four are links of the original square lattice along directions $1~\& ~2$, and remaining two are virtual links, resulted from point splitting along direction $3$.  The electric fields are  defined at each end of the links of the original square lattice. Hence even for hexagonal lattice, only the electric fields for links along directions $1~\& ~2$ contribute to the Hamiltonian given in (\ref{ham}). In prepotential formulation, this electric part of Hamiltonian $H_E$ counts the fluxes which are actually related to number of prepotentials sitting at each end of the  links of the original lattice \cite{pp}.
 
 The magnetic part of the Hamiltonian $H_{mag}$ is anyway more complicated to analyse. Clearly, at weak coupling regime, this part contributes most, and hence it is essential to simplify it as much as possible to make analytic calculations feasible. The part however is altered from the square lattice, as it must contain trace of the product of link operators along the full hexagonal plaquette to make the smallest closed loop. 
 
%

The magnetic Hamiltonian on hexagonal lattice contains $2^6$ different gauge invariant plaquette terms for prepotential formulation of SU(2) gauge theory on hexagonal lattice. Pictorially, these plaquette terms contain $n_d$ number of dotted link and $6-n_d$ number of solid links around a plaquette, for $n_d=1,2,..,6$. These solid and dashed line comes in different combination yielding all of the $64$ plaquette terms as illustrated in figure \ref{plaq64detail}. 
 
\begin{figure}[t]
\begin{center}
\includegraphics[width=0.3\textwidth,height=0.4\textwidth]
{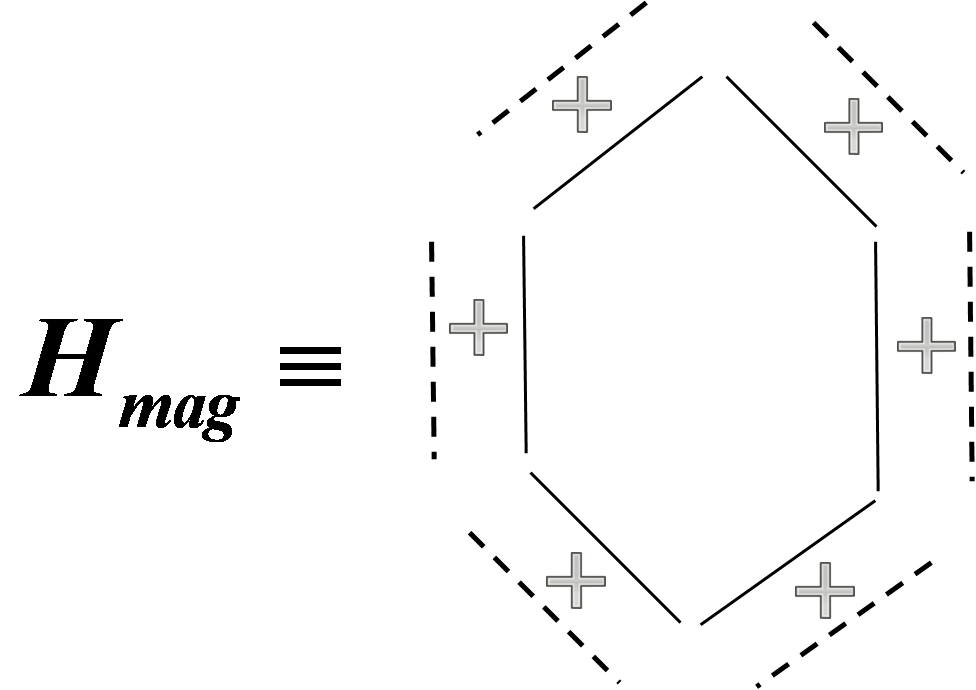}
\end{center}
\caption{The hexagonal plaquette consisting of six links each of which can be either a solid line or a dashed line within prepotential framework. Hence all possible plaquette operators contributing to the magnetic part of the Hamiltonian consists of $2^6=64$ plaquette diagrams with a combination of solid and dashed lines around each links. } 
\label{64plaq}
\end{figure}

Let us now, concentrate on the loop configurations of the system and proceed towards a mean field ansatz (i.e average loop configuration) for low energy limit of the theory.

We have already characterized the local loop states at a particular site $x$ of a hexagonal lattice in terms of linking numbers as $|l_{12},l_{23},l_{31}\rangle_x$ in (\ref{genloopst}) and also discussed how these linking numbers are related to the occupational number basis in (\ref{l2n}). The original Wilson loops, which are non-local are obtained by weaving these local loops at neighbouring sites along the three directions, using an extra Abelian Gauss law \cite{pp} constraint at each link direction  as given below:
\bea
\label{u1gl}
n_i(x)=n_i(x+e_i) ~~\mbox{for } i=1,2,3.
\eea

\subsection{Mean Field Ansatz} 

In this sub-section, we make an ansatz for the  vacuum loop configuration  of the SU(2) lattice gauge theory. Strong coupling vacuum of the system is well-known and consists of $0$ flux state. Whereas, in the naive continuum limit, as $g\rightarrow 0$, all the loop configurations contribute to the low energy spectrum. However, the maximum contribution is expected to come from large loops carrying large fluxes. In prepotential formulation, the size of the loop is not relevant, as all the loops has been made local. Abelian weaving along the links give rise to the standard Wilson loops. 

In this context, let us make a general ansatz for the vacuum loop configuration of the system irrespective of the coupling regime.  Let us assume that the low energy loop configurations are given by same amount of  fluxes flowing across each site.
Note that, we are working on a virtual hexagonal lattice in two dimensions as discussed before. Note that, on these lattices, there exists two different types of flux, flowing across each site. The standard flux flowing along the direction of original links, i.e between $\{12\}$ direction is the real flux. Moreover, there exist the fluxes, which flows from a real direction to a virtual direction, namely $\{23\}$ and $\{31\}$. We consider this, and make an ansatz that,
\bea
\label{ansatz}
l_{12}(x)=L~; ~l_{23}(x)=M~;~ l_{31}(x)=M~ \forall x
\eea
where, $l_{ij}(x)$'s are the linking quantum number (measuring fluxes flowing across $\{ij\}$ directions) specifying the loop states in (\ref{genloopst}) at a site $x$. 
 The electric part of the Hamiltonian (\ref{ham}) in Prepotential formulation reads as:
\bea
H_e=g^2 \sum_{\mbox{links}} E^{2}_{links}= g^2 \sum_x \left[ \frac{n_1(x)}{2}\left(\frac{n_1(x)}{2}+1\right) +  \frac{n_2(x)}{2}\left(\frac{n_2(x)}{2}+1\right) \right]
\eea
where, $n_1$ and $n_2$ counts the number of prepotentials residing on links along $1\& 2$ directions respectively. This occupation numbers are related to the linking numbers at each site x and hence with mean values $L \&M $  as,
\bea
n_1=l_{12}+l_{31} &=& L+M ~~~~ \\
n_2=l_{12}+l_{23} &=& L+M ~~~~\mbox{using (\ref{ansatz})}
\eea
Implying,
\bea
H_e= g^2 \sum_x  \frac{1}{2}(L+M)(L+M+2).
\eea
Next we concentrate on the magnetic part of the Hamiltonian within this mean field approximation. The magnetic Hamiltonian contains 64 plaquette operators given in figure \ref{plaq64detail}.
Each of these 64 plaquette terms consists of 6 local loop operators given in figure \ref{hexloopop}, which comes with a coefficients listed in table \ref{tablecoeff} which are functions of number operators (i.e linking numbers or occupation numbers). However,  within the mean field ansatz they can be regarded as C-numbers (i.e functions of constants L and M) as listed in \ref{coeff_list_mf}.
\begin{table}
\label{coeff_list_mf}
\begin{center}
\begin{tabular}{|c|c|}
\hline
Loop action & \mbox Explicit action on $|L,M\rangle$  \\  
\hline  \includegraphics[width=0.08\textwidth, height=8mm]{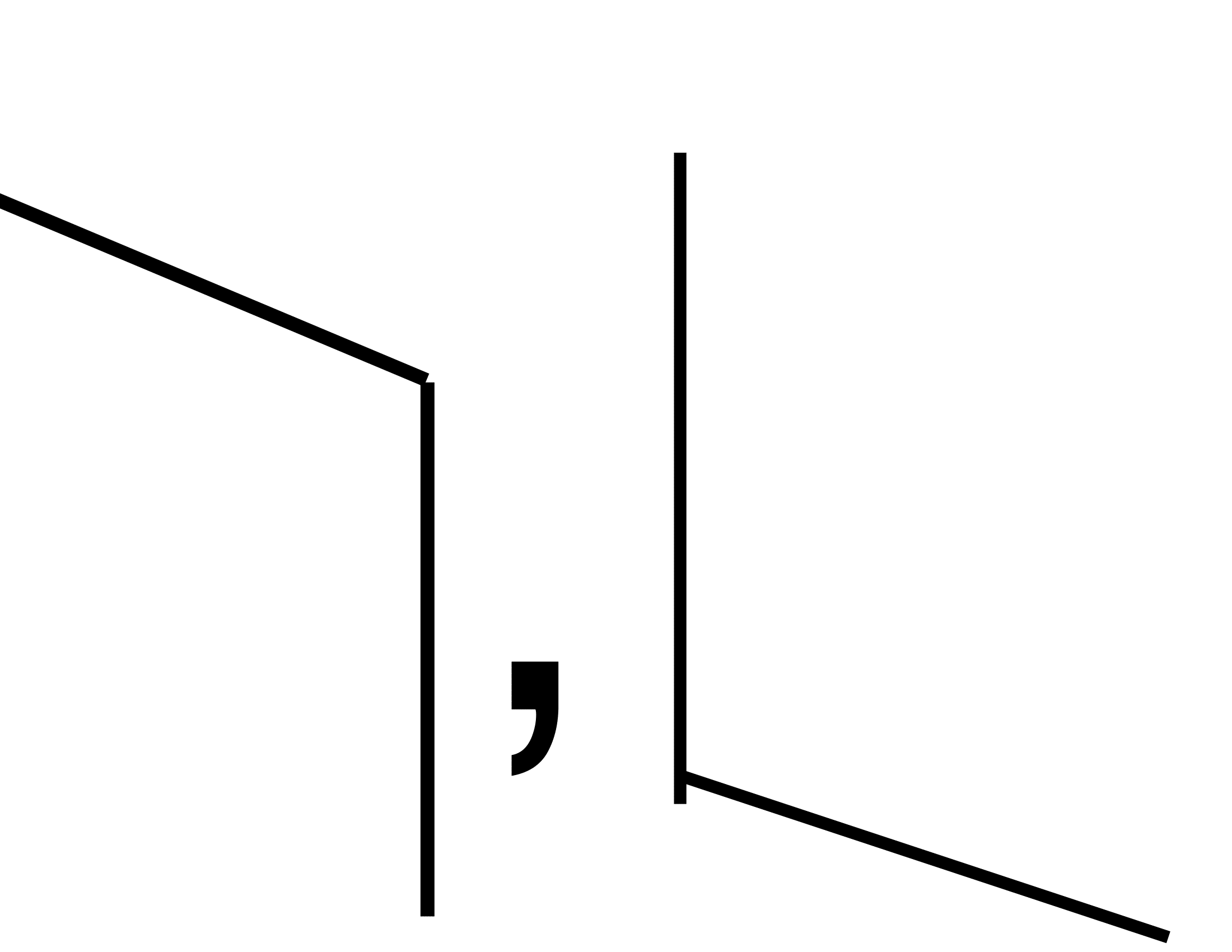} & $\sqrt{\frac{(M+1) (L+2 M+2)}{(2 M+1) (L+M+2)}}|M+1\rangle$  \\
\hline  \includegraphics[width=0.12\textwidth, height=8mm]{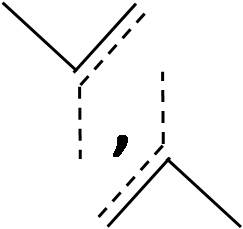}& $-\sqrt{\frac{(L+1) M}{(2 M+2) (L+M+1)}}|L+1,M-1\rangle$\\
\hline   \includegraphics[width=0.12\textwidth, height=8mm]{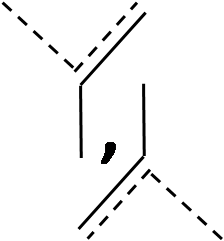} & $-\sqrt{\frac{L (M+1)}{(2 M+1) (L+M+2)}}|L-1,M+1\rangle$ \\
\hline \includegraphics[width=0.12\textwidth, height=8mm]{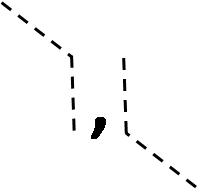} & $\sqrt{\frac{M (L+2 M+1)}{(2 M+1) (L+M+2)}}|M-1\rangle$  \\
\hline \includegraphics[width=0.12\textwidth, height=8mm]{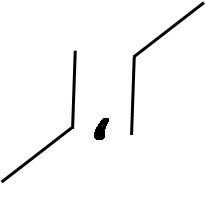} & $\sqrt{\frac{(M+1) (L+2 M+2)}{(2 M+2) (L+M+1)}}|M+1\rangle$ \\
\hline  \includegraphics[width=0.12\textwidth, height=8mm]{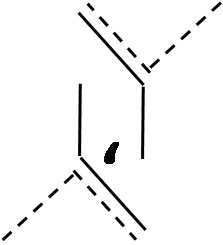} & $-\sqrt{\frac{L (M+1)}{(2 M+1) (L+M+2)}}|L-1,M+1\rangle$ \\
\hline  \includegraphics[width=0.12\textwidth, height=8mm]{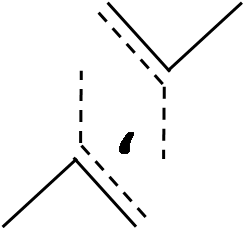} & $-\sqrt{\frac{(L+1) M}{(2 M+2) (L+M+1)}}|L+1,M-1\rangle$ \\
\hline  \includegraphics[width=0.12\textwidth, height=8mm]{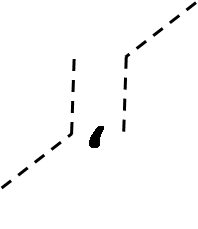} & $\sqrt{\frac{M (L+2 M+1)}{(2 M+2) (L+M+1)}}|M-1\rangle$ \\
\hline \includegraphics[width=0.12\textwidth, height=8mm]{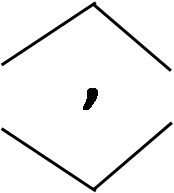} & $\sqrt{\frac{(L+1) (L+2 M+2)}{(L+M+2) (L+M+1)}}|L+1\rangle$ \\
\hline  \includegraphics[width=0.12\textwidth, height=8mm]{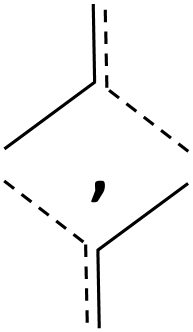} & $-\sqrt{\frac{(M+1) M}{(L+M+2) (L+M+1)}}|L,M\rangle$ \\
\hline  \includegraphics[width=0.12\textwidth, height=8mm]{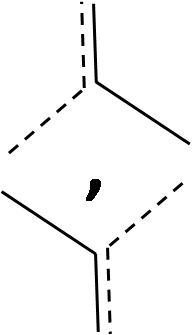} & $-\sqrt{\frac{(M+1) M}{(L+M+2) (L+M+1)}}|L,M\rangle$ \\
\hline \includegraphics[width=0.12\textwidth, height=8mm]{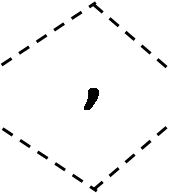} & $\sqrt{\frac{L (L+2 M+1)}{(L+M+2) (L+M+1)}}|L-1\rangle$ \\
\hline
\end{tabular}
\end{center}
\caption{List of coefficients under the mean field ansatz}
 \end{table}
Combining these coefficients we get $2^6$ coefficients seating in front of each of the Magnetic term given in figure \ref{64plaq}, which we call as $\mathcal C_i$, for $i=1$ to $64$.

\begin{figure}[t]
\begin{center}
\begin{overpic}[scale=0.4,unit=5mm]{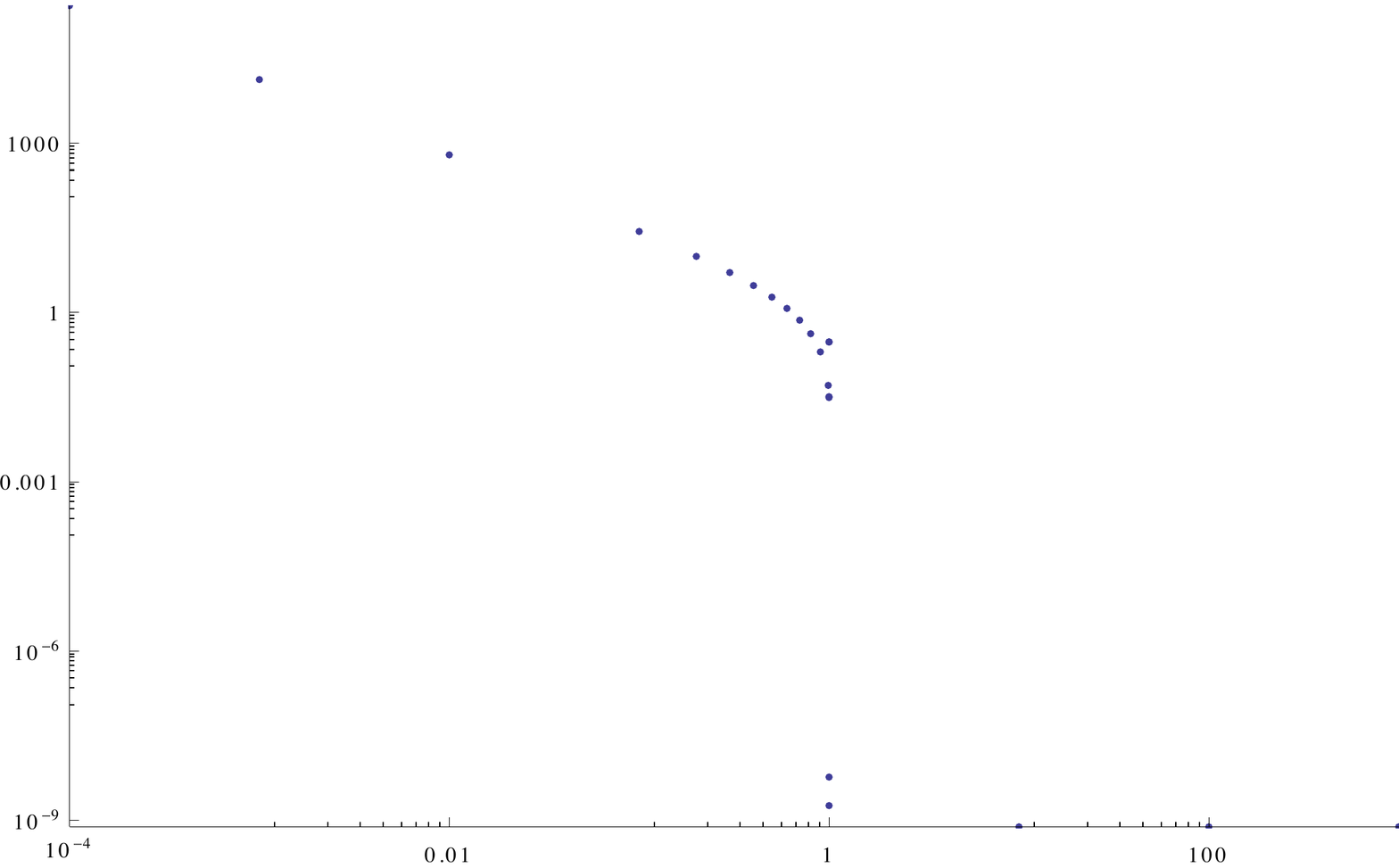}
\put(-2,13){$L|_{H_{min}}$}
\put(11.3,5){$g$}
\end{overpic}
\vspace{-8mm}
\caption{The numerical value of the mean field variable L (M) for which the Hamiltonian function reaches a minima , i.e $L|_{H_{min}}$ is plotted against the coupling g. The curve shows a clear jump in the average loop configuration, moving from strong to weak coupling regime of the theory. } 
\label{pt}
\end{center}
\end{figure}

Now, in order to fix the numerical values of $L$ and $M$, in our analysis, we calculate the Hamiltonian for the limiting case, when the all the plaquette operators (except the coefficients) becomes $1$. Now, the Hamiltonian function for each plaquette reads as:
\bea
H(g,L,M)= g^2  \frac{1}{2}(L+M)(L+M+2)+ \frac{1}{g^2}(4-\sum_{i=1}^{64}\mathcal C_i)
\eea
We now minimize this function for different values of the coupling $g$. This minimization yields a set of L and M values at different couplings, for which the Hamiltonian function reaches a minima. We  plot this set of $L$ values at $H_{min}$ in figure \ref{pt} to get a clear notion of phase transition between the mean field phases at weak and strong coupling regime occurring exactly at $g=1$.  A very similar curve is obtained for $M$ as well, which shows the exact same nature except the fact that for each values of $g$, the numerical value of $M$ is much less than the corresponding $L$ in weak coupling regime, i.e $L>>M>>0$. This same result will be again obtained in a more precise calculation for the proposed reduced Hamiltonian in the next section. 

Finding the spectrum for  the full Kogut-Susskind Hamiltonian even in the mean field approximation is still a challenge at this stage. In the next section we construct a reduced Hamiltonian which describes the dynamics within the Hilbert space of the above mentioned average loop configurations.

\section{The reduced Hamiltonian and its spectrum}

In the last section we have made a mean field ansatz, in which an arbitrary loop state defined at each site of the virtual hexagonal lattice is given by the average value for the linking number variables is given by:
\bea
|L,M,M\rangle_x~~~~\forall x
\eea

In this section, we will  reduce the  full Kogut-Susskind Hamiltonian  to a sub Hamiltonian, which describes the dynamics within this mean field ansatz. With this reduced system, one can get reasonable physical results with minimal calculational effort and hence establishes this as a valid toy system to understand weak coupling regime of the gauge theory analytically, numerically as well as by quantum simulating the system.
\subsection{The Sub-Hamiltonian}

As we have already explained in the last section, the electric part of the Kogut-Susskind Hamiltonian in the mean field approximation reads as:
\bea
H_e\equiv  g^2 E^2 = g^2 \sum_x  \frac{1}{2}(L+M)(L+M+2).
\eea

Next we concentrate on the magnetic part of the Hamiltonian within this mean field approximation. The magnetic Hamiltonian contains 64 plaquette operators, each of which is a combination of six local vertex operators listed in table \ref{coeff_list_mf} within the mean field ansatz.
 \begin{figure}[h]
\begin{center}
\includegraphics[width=0.6\textwidth,height=0.6\textwidth]
{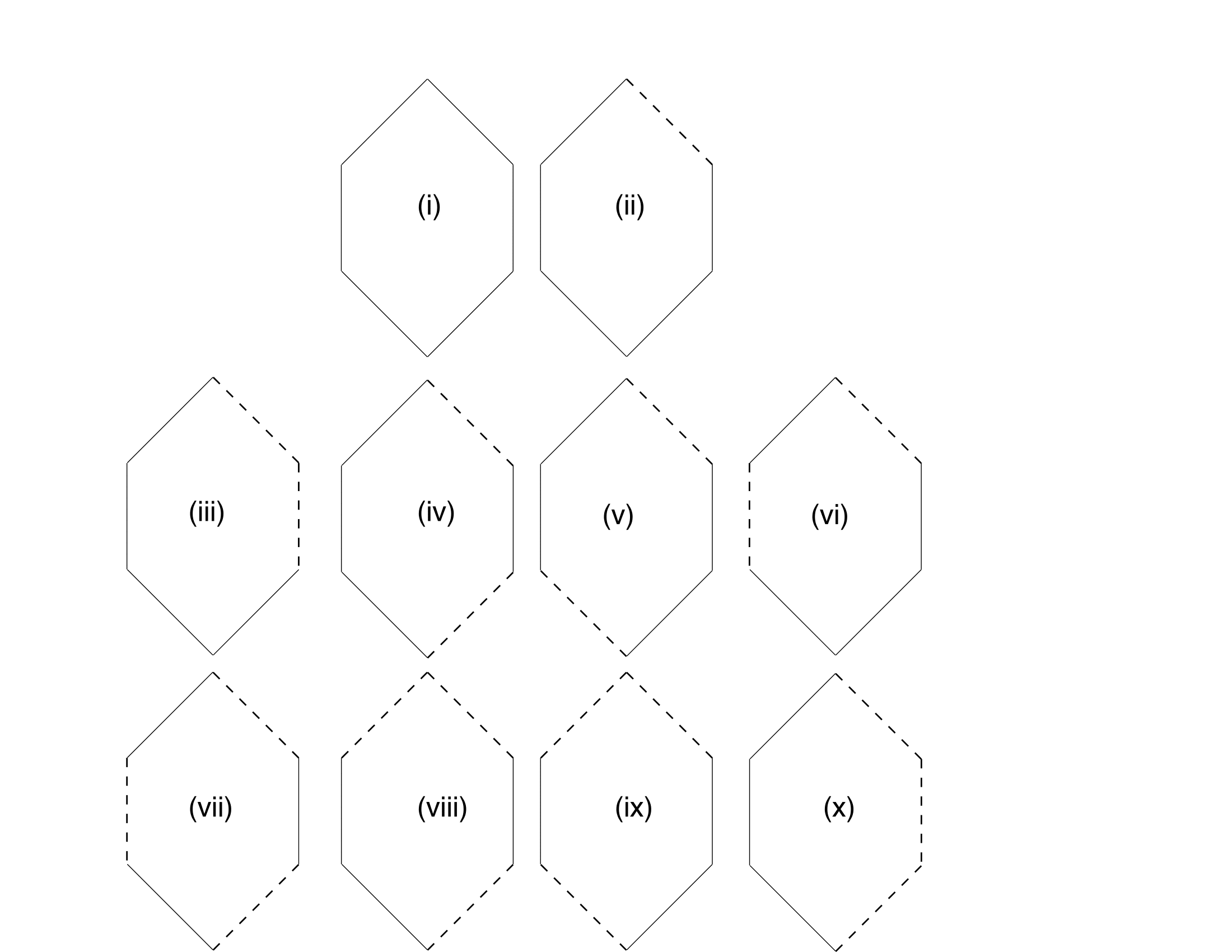}
\end{center}
\caption{Pictorial representation of plaquette operators appearing in the magnetic Hamiltonian on hexagonal lattice. All of these terms along with their rotationally symmetric and hermitian conjugate pairs constitute the 64 plaquette terms.
} 
\label{plaq64detail}
\end{figure}
 Let us now analyze each and every plaquette term illustrated in figure \ref{plaq64detail} :\\
(i) This term together with its Hermitian conjugate pair (2 terms) creates or annihilates flux around a full plaquette.
 (ii) This term along with its rotationally symmetric (6 terms) and hermitian conjugate pairs (6 more terms) increases or decreases the length of the Wilson loops by 5 units.
 (iii) Same as (ii), total 12 terms, changes length by 4 units. (iv), (v), (vi) and the rotationally symmetric 9 terms ( 9 more hermitian conjugate terms) merge (separate) two loops and construct one (two) bigger (smaller) loop(s) of their combined length $+(-)$ 4 units of length. (vii) and its hermitian conjugate term (total 2 terms) are rotationally symmetric and merges three loops to construct e bigger loop of their combined length. (viii), (ix) and their rotationally symmetric 6 terms for each constitute a hermitian conjugate set of 12 terms, each of which merges two loops and construct a bigger one of the same length. Finally (x) and its 6 rotationally symmetric terms are hermitian conjugate set and changes shape of a loop without changing its length. 
 

Let us now make the following observations:\\
Among the 64 plaquette terms given in figure
, there are certain terms which create or annihilate net fluxes around a plaquette. These plaquette terms can indeed build up the complete loop space, starting from strong coupling vacuum.  However, we are interested in a particular state of the system, where each of the loops takes an average value throughout the lattice, hence there really exists no scope for creating or annihilating any net flux around any of the plaquettes. Hence, we choose only a subset of these 64 plaquette terms which keeps the dynamics within the mean field ansatz. More specifically, we choose a sub-magnetic Hamiltonian, which does not create or annihilate any net flux around a plaquette, but rearranges the internal loop configurations.  Note that, (vii),(viii),(ix) and (x) type terms in figure \ref{plaq64detail} are loop operators that do not change any net flux flowing across the lattice, but just rearranges the intertwiners across different sites. These plaquette operators form a rotationally symmetric hermitian operatorThose are the terms which has equal number of creation and annihilation of links around a plaquette, pictorially represented by a plaquette with 3 solid link and 3 dotted link.   These special terms are explicitly given as below:
 \begin{enumerate}
 \item Plaquette consisting of alternate solid and dashed line. There are only  two options for this, which constitutes the rotationally symmetric and Hermitian operator. We denote this operator by
 \bea
 H_{\mbox{pmpmpm}}+\mbox{rotation}
 \eea
 \item Plaquette consisting of three consecutive solid and dashed lines. There are six options for this, which constitutes the rotationally symmetric and Hermitian operator. We denote this operator by
 \bea
 H_{\mbox{pppmmm}}+\mbox{rotations}
 \eea
  \item Plaquette consisting of two consecutive solid line, two consecutive dashed line and then a single solid and dashed line along with their Hermitian conjugate plaquette terms.   We denote this type of operator by
 \bea
 H_{\mbox{ppmmpm}}+H_{\mbox{mmppmp}}+\mbox{rotations}
 \eea
Each of these two types of plaquette terms has 6 rotationally symmetric contribution. Hence this particular type of rotational symmetric Hermitian operator contains total of 12 individual plaquette operators.
 \end{enumerate}
 Combining the above three, we find that the magnetic part of the reduced Hamiltonian contains $2+6+12=20$ plaquette operators and is completely rotationally symmetric and Hermitian. 

Let us now construct a particular basis consisting of the state $|L,M\rangle$ and the twenty plaquette operators listed above, acting on it. Hence we have a $21$ dimensional basis vectors listed as below:

\bea
\label{21states}
&&|L,M\rangle ~~; ~~ \{H_{\mbox{pmpmpm}}|L,M\rangle\} ~~;~~\{H_{\mbox{pppmmm}}|L,M\rangle\}~~;~~ \{ H_{\mbox{ppmmpm}}|L,M\rangle\}
\eea
where, the braces in the last three states denote the sets consisting of rotationally symmetric and Hermitian conjugate states. Each of these states consists of one of those three types of vertices at each site. We first calculate the local action of $E^2$ on each vertex, depending on whether it is a junction of two solid line (pp-vertex $\equiv |pp\rangle$), two dashed line (mm-vertex $\equiv |mm\rangle$) and of one solid and one dashed line (pm vertex $\equiv |pp\rangle$). 
 Using the 12 vertices described in table \ref{coeff_list_mf}, the action of $E^2$ within mean field approach is given by:
\bea
E^2|pp\rangle &=& \frac{1}{2}(L+M+1)(L+M+3)|pp\rangle \equiv v_p|pp\rangle\nonumber \\
E^2|mm\rangle &=& \frac{1}{2}(L+M-1)(L+M+1)|mm\rangle\equiv v_m |mm\rangle \label{vpvm} \\
E^2|pm\rangle &=& \frac{1}{2}(L+M)(L+M+2)|pm\rangle\equiv v|pm\rangle \nonumber \\
\eea
However, our basis now consists of  21 states given in (\ref{21states}), each having six of the above vertices. The action of $E^2$ on these states are given as below:
\bea
 E^2|L,M\rangle &=& 6v|L,M\rangle \nonumber \\
E^2\left[H_{\mbox{pmpmpm}}|L,M\rangle\right] &=& 6v\left[H_{\mbox{pmpmpm}}|L,M\rangle\right]\nonumber \\
E^2\left[H_{\mbox{pppmmm}}|L,M\rangle\right] &=&(2v+2v_p+2v_m)\left[H_{\mbox{pppmmm}}|L,M\rangle\right]\nonumber \\
E^2\left[H_{\mbox{ppmmpm}}|L,M\rangle\right] &=&(4v+v_p+v_m)\left[H_{\mbox{pppmmm}}|L,M\rangle\right]\label{ei}
\eea

At this point, we import another notation for the basis. Let the basis be denoted as $|i\rangle, for i=0,..,20$, with $|0\rangle=|L,M\rangle$, and $|i\rangle=H_i|0\rangle$ for $i=1,2,..,20$, where $H_i$ denotes the $20$ plaquette operators described before (given as $H_{\mbox{pmpmpm}}, H_{\mbox{pppmmm}}, H_{\mbox{ppmmpm}}$). Hence, the reduced Hamiltonian takes the form:
\bea
H_{\mbox{reduced}}=\sum_{plaquettes}\left[ g^2 E^2+\frac{1}{g^2}(1-\sum_{i=1}^{20} H_i)\right]
\eea
 The electric part of the Hamiltonian acting on a state of this basis gives:
\bea
g^2E^2|i\rangle=E_i|i\rangle,   ~~~~ i=0,1,..,21
\eea
and the magnetic part gives:
\bea
\frac{1}{g^2}(1-\sum_{i=1}^{20}H_i)|i\rangle=\frac{1}{g^2}|i\rangle-\sum_{i=1}^{20}\frac{\mathcal C_i}{g^2}|0\rangle
\eea
In this particular basis, the Hamiltonian matrix for each plaquette takes the form given below:
\bea
\left(\begin{array}{ccccccc}
g^2E_0+1/g^2 & -\frac{\mathcal C_1}{g^2} &-\frac{\mathcal C_2}{g^2} & \ldots & -\frac{\mathcal C_{20}}{g^2} \\ 
-\frac{\mathcal C_1}{g^2} &g^2E_1+1/g^2 & 0 & 0 & 0 \\ 
-\frac{\mathcal C_2}{g^2} & 0 & g^2E_2+1/g^2 & 0 & 0 \\ 
\vdots & \vdots & \vdots & \ddots & \vdots  \\ 
-\frac{\mathcal C_{20}}{g^2}& 0 & 0 & 0 & g^2E_{20}+1/g^2
\end{array} \right)
\eea
Note that, the diagonals have contribution from the electric term as well from the constant term in magnetic Hamiltonian. Other than the diagonal entries, other non zero matrix elements of the Hamiltonian lies along the first row and first column of the matrix, and those are basically given by the coefficients sitting in front of the 20 plaquette terms discussed above. We label these coefficients as $\mathcal C_i$, for $i=1,..,20.$ The advantage of working with such a basis is that the Hamiltonian matrix takes a very special form, namely form of an arrowhead matrix which in turn enables us to solve for the eigenvalues  analytically.

Let us assume the Hamiltonian matrix to satisfy the following eigenvalue equation:
\bea
\left(\begin{array}{ccccccc}
g^2E_0+1/g^2 & -\frac{\mathcal C_1}{g^2} &-\frac{\mathcal C_2}{g^2} & \ldots & -\frac{\mathcal C_{20}}{g^2} \\ 
-\frac{\mathcal C_1}{g^2} &g^2E_1+1/g^2 & 0 & 0 & 0 \\ 
-\frac{\mathcal C_2}{g^2} & 0 & g^2E_2+1/g^2 & 0 & 0 \\ 
\vdots & \vdots & \vdots & \ddots & \vdots \\ 
-\frac{\mathcal C_{20}}{g^2}& 0 & 0 & 0 & g^2E_{20}+1/g^2
\end{array} \right)
\left( \begin{array}{c}
a_0\\a_1\\a_2\\\vdots \\ a_{20}
\end{array} \right)
=\lambda \left( \begin{array}{c}
a_0\\a_1\\a_2\\\vdots \\ a_{20}
\end{array} \right)
\eea
This matrix equation is equivalent to the set of eigenvalue equations mentioned below:
\bea
\left( g^2E_0+\frac{1}{g^2} \right)a_0-\frac{1}{g^2}\sum_i \mathcal C_ia_i &=& \lambda a_0 \label{a0} \\
-\frac{\mathcal C_i}{g^2}a_0+\left( g^2E_i+\frac{1}{g^2} \right)a_i &=& \lambda a_i \label{ai}
\eea
Now, from (\ref{ai}),
\bea
\left( g^2E_i+\frac{1}{g^2}-\lambda \right)a_i &=& \frac{\mathcal C_i}{g^2}a_0 \nonumber \\
\Rightarrow a_i&=&\frac{\mathcal C_i a_0}{1-g^2\lambda+g^4E_i} \label{aia0}
\eea
Putting (\ref{aia0}) back in (\ref{a0}), we get,
\bea
\left( g^4E_0+1 \right)a_0-\sum_i \frac{\mathcal C_i^2 a_0}{1-g^2\lambda+g^4E_i}&=& g^2\lambda a_0 \nonumber \\
\Rightarrow \left( g^4E_0+1 \right)a_0-\sum_i \frac{\mathcal C_i^2 a_0}{1-g^2\lambda+g^4E_i}&=& g^2\lambda a_0 \nonumber \\
\Rightarrow \tilde{\lambda} = \sum_i \frac{C_i^2}{\tilde{\lambda}-g^4(E_0-E_i)}, &&~\mbox{where, } \tilde{\lambda}=1-g^2\lambda+g^4E_i \label{lambdaeq}
\eea
Solving (\ref{lambdaeq}) would yield the $\tilde{\lambda}$, which in tern gives all the eigenvalues of the $21\times 21$ Hamiltonian matrix. To simplify (\ref{lambdaeq}), we further make the following observation. Note that, the diagonal elements $g^2E_i+1/g^2$ are highly degenerate, as one finds in (\ref{ei}) that, $E_i$'s can take only three possible values, namely $6v, ~(2v+2v_p+2v_m)$ and $(4v+v_p+v_m)$. Hence, the sum in (\ref{lambdaeq}) reduces to the following sum of three terms:
\bea
\label{lambdat}
\tilde{\lambda}=\frac{\tilde{\mathcal C_1}}{\tilde{\lambda}+2g^4}+\frac{\tilde{\mathcal C_2}}{\tilde{\lambda}+g^4}+\frac{\tilde{\mathcal C_3}}{\tilde{\lambda}}
\eea
where, $\tilde{\mathcal C_1}, ~\tilde{\mathcal C_2},~ \tilde{\mathcal C_3}$ are combinations of the  coefficients ${\mathcal C_i}$ taking into account of the degeneracy.
(\ref{lambdat}) is a transcendental equation and can be solved graphically, however approaching a rough solution is quite easy. We plot the right hand side of this equation and get divergences at three poles precisely at $\tilde{\lambda}=0, g^4, 2g^4$ respectively. The plot of left hand side, i.e straight line $y=x$ cuts the right hand side curve very close to the position of the poles. The exact value of $\tilde{\lambda}$ yields the value of the eigenvalues as:
\bea
\label{lambda}
\lambda=\frac{(1-\tilde{\lambda}+g^4*E_0)}{g^2}
\eea
Note that, the smallest eigenvalue $\lambda$, 
actually corresponds to the highest $\tilde{\lambda}$ from 
(\ref{lambda}).
To get the exact numerical value of the solution one needs to consider the coefficients $\tilde{\mathcal C_i}$'s appearing in right hand side of
(\ref{lambdat}). Note that, the coefficients $\tilde{\mathcal C_i}$'s are functions of the mean fields $L$ $\&$ $M$.
Hence, it is necassary to calculate the exact value of these mean fields. For this purpose, we take the following approach. 
\begin{table}
\begin{center}
\begin{tabular}{|c|c|c|}
\hline 
$g$  & $L|_{\lambda_{min}}$ &$M|_{\lambda_{min}}$\\ 
\hline 
10& 0&0 \\
\hline
1 & 0&0 \\
\hline
0.1 & 10 & 3 \\ 
\hline 
0.01 & 374 & 32 \\ 
\hline 
0.001 & 11620 & 320 \\ 
\hline 
\end{tabular} 
\label{LMdata}
\caption{The smallest eigenvalue of the Hamiltonian matrix reaches its minima for these values of $L,M$, at the particular $g$ mentioned in the first column.}
\end{center}
\end{table}
\begin{itemize}
\item For 
some fixed $g$, we numerically calculate (using Mathematica) the value of $L$ and $M$, for which lowest eigenvalue of the Hamiltonian matrix reaches a minima. In the strong coupling regime, i.e for $g\geq 1$, that minima 
is always at $L=M=0$. However, for smaller and smaller values of $g$, the 
minima is at larger and larger values of $L\&M$, as listed in table
4.   This calculation establishes the naive analysis done in section 4 of this paper which shows the existence of two different mean field phases of the system in strong and weak coupling regimes. 
Note that, we are interested in the weak coupling regime of the theory as the continuum limit lies there. Upto this point of this work, we have not used any assumption for weak coupling limit, except taking a mean field ansatz. We now fix the mean field configuration in a way, such that we are in the weak coupling regime of the theory. From the variational study discussed above, we see, that the ground state energy shows a perfect first order phase transition at coupling $g=1$, above which the vacuum is the strong coupling vacuum, which in the mean field ansatz gives $|L=0,M=0\rangle$. However, as $g\rightarrow 0$, the lowest energy mean field state turns out to be a state comprising of large average flux at each site, implying the weak coupling vacuum to be consisted of loops carrying large fluxes, throughout the lattice. As $g\rightarrow 0$, $L>>M>>0$. 
The coefficients, listed in table 3 also changes to particular limiting values as $g\rightarrow 0$.
\item
We now consider a particlular value of $g$, in the weak coupling regime, and the corresponding mean field configuration. For each configuration, we exactly diagonalize the Hamiltonian matrix, and calculate the eigenvalues. 
We list the spacings between the lowest one and first excited one as $\Delta \lambda_1$ and similarly between the first and second excited one as $\Delta \lambda_2$. We list these two gaps and their ratios at different values of the coupling constant $g\rightarrow 0$ in table \ref{gap}. 
\begin{table}
\label{gap}
\begin{center}
\begin{tabular}{|c|c|c|c|}
\hline 
$g$  & $\Delta\lambda_1$ &$\Delta\lambda_2$ & $\frac{\Delta\lambda_1}{\Delta\lambda_2}$\\ 
\hline 
0.1 & 0.00998711 & 0.00916295 & 1.08994\\ 
\hline 
0.01 & 0.0000999971 & 0.0000989114 & 1.01098 \\ 
\hline 
0.001 & 0.00000099896 & 0.000000998378 & 1.00058 \\ 
\hline 
\end{tabular} 
\caption{The gaps between three consecutive energy levels, and their ratios are listed for different values of coupling $g$ in weak coupling regime.}
\end{center}
\end{table}
It is very much clear from table \ref{gap}, that the gaps are scaling as $\sim g^2$ and the ratio of lowest two energy gaps converges to the numerical value of 1. Note that, the absolute value of masses are subjected to be renormalized. However, the ratio of the two consecutive mass-gaps are always physical. In our study we have shown it to converge to numerical value of 1, for $g\rightarrow 0$ for arbitrarily large lattices, as all the computation was done locally at each site.
\end{itemize}
The scaling of mass-gaps $\sim g^2$ is consistent with both the weak coupling perturbation expansion in  \cite{muller-ruhl}\footnote{$\Delta E=0.2637g^2$}, variational analysis \cite{arisue, Guo}\footnote{$\Delta E\approx 2g^2$}, using cluster algorithm \cite{cluster} \footnote{$\Delta E\approx 2.2g^2$} for SU(2) gauge theory on $2+1$ dimensional lattice as well as with Monte Carlo study on $3$dimensional Euclidean lattice \cite{mc}\footnote{$\Delta E\approx 2.1 g^2$}.
 

On a further note, the analytically calculated value of the ratio of energy differences for the reduced Hamiltonian within the mean field ansatz, matches with previous calculations at weak coupling limit for finite lattices. For example, the mass spectrum of $0^{++}$ sector for SU(2) lattice gauge theory in $2+1$ dimensions is obtained in \cite{arisue} as follows: 
 \bea
&&m_{1}a=(2.056±0.00l)g^2~,~m_2a=(3.64 ±0.03)g^2~,~m_3a=(5.15±0.10)g^2 \nonumber \\
\Rightarrow && \Delta \lambda_1\approx 1.58g^2 ~\&~ \Delta \lambda_2\approx 1.51g^2 \nonumber \\ \Rightarrow && ~~~~\frac{\Delta \lambda_1}{\Delta \lambda_1}=1.046
\label{matching}
\eea
This result is obtained with a lattice consisting of 25 plaquettes and in the weak coupling region $1.8<1/g^2<3.6$. Note that, the results obtained in \cite{arisue} are consistent with the results obtained with Monte Carlo calculations within Euclidean formalism as well.

(\ref{matching}) establishes the validity of our much simplified system to extract out weak coupling results for practical purposes, which can be exploited to address various problems in future researches.

\section{Summary and future directions}

In this work we have proposed and justified an effective mean field description for the low energy spectrum of SU(2) lattice gauge theory in $2+1$ dimension and have calculated the spectrum at the weak coupling limit  analytically. Starting from prepotential formulation on square lattice, we perform virtual splitting of each lattice site into two and end up with a virtual hexagonal lattice. On this hexagonal lattice, all of the local loop states in prepotential formulations constitutes an exact and orthonormal loop basis, with no further Mandelstam constraints. We have proposed a mean value ansatz  for the loop configurations throughout the lattice contributing to the low energy spectrum of the theory. We have shown that such average loop configurations have two distinct phases at the strong and weak coupling regime. Next, we have chosen a reduced Hamiltonian, from the full Kogut-Susskind Hamiltonian, which keeps the dynamics of the loops confined into our ansatz. Variational study shows that this reduced system with mean value-loop configuration shows a clear jump between the weak and strong coupling vacuum. As we are interested to explore weak coupling regime of the theory, we choose the relevant average loop configuration in that regime and calculate the spectrum for the reduced Hamiltonian we choose. In this spectrum we find  $\Delta E\sim g^2$ which is the expected weak coupling behaviour for mass gap of the theory. The spacings of the spectrum obtained in this work is as well consistent with the available literature at weak coupling regime of $2+1$ dimensional SU(2) lattice gauge theory. We have discussed in detail, how the point splitting lattices are constructed in higher dimensions, which can as well be exploited to extend this work beyond $2+1$ dimensions in a straight forward way. 

In a recent and parallel work \cite{new}, the point splitting lattice is constructed and utilized to analytically study the weak coupling limit of SU(2) lattice gauge theory in $2+1$ dimension as well. In that work, they have used the path integral representation of the phase space to analytically compute the dispersion relation at the lowest order in weak coupling perturbation expansion. 

However, the particular study demonstrated in this paper shows that the physical results at the weak coupling regime of SU(2) gauge theory can be extracted from a much simpler mean field approximation made within prepotential formulation of the theory. Being completely gauge invariant, and formulated  only in terms of relevant physical degrees of freedom, this technique is suited for both analytic calculations and numerical simulations. From analytic perspective, this study gives a clear notion of the weak coupling vacuum for pure gauge theory and its dynamics. From numerical perspective, this particular formulation is most suited for quantum Monte-Carlo simulation of Hamiltonian lattice gauge theory using a complete gauge invariant basis characterized by only integers. Till date, this aspect has not been studied extensively, but worth investigating in near future. That study will lead to explore some of the very important physics such as calculation of the entanglement entropy of lattice gauge theory. Last but not the least, there is a tremendous progress going on, in the recently developed research interests for quantum simulating gauge theories using ultracold atoms in optical lattices as well as using different other techniques. The prepotential formulation has already been explored to propose quantum simulator for gauge theories \cite{qs}. However, most of such proposals till date have addressed the abelian gauge theories and the strong coupling regime. This present work, shows the way to construct quantum simuator to simulate the loop dynamics of non-Abelian lattice gauge theory at weak coupling regime using ultracold atoms in optical lattices. The work in this direction is in progress and will be reported shortly.
 
\section*{Acknowledgement}
We would like to thank Ramesh Anishetty for numerous discussions throughout the entire project as well as  for his valuable suggestions on the manuscript. We would also like to thank Pushan Majumdar for useful  discussions and Department of Theoretical Physics, Indian association for the cultivation of science, Kolkata for support during a substantial part of the project.

\appendix
\section{Orthonormal loop basis and its dynamics on  square lattice}

Within prepotential framework of the theory one is equipped with a set of orthonormal basis states, defined at each lattice site \cite{pp,prd}. 
For SU(2) lattice gauge theory on 2 dimensional spatial lattice, such an orthonormal basis can be easily obtained in terms of angular momentum fluxes. In this section, we briefly quote the result of plaquette (the smallest closed loop) dynamics from \cite{pp} which we have used to compare with the dynamics of hexagonal plaquette.

At each site of a two dimensional square lattice, four links are attatched in directions $1,2,\bar 1~ \& ~ \bar 2$  carrying the angular momentum fluxes $j_1,j_2,j_{\bar 1} ~\& ~ j_{\bar 2}$ respectively ($j_i=n_i/2$). The gauge invariant state at a site $x$ must have the net angular momentum, i.e the sum of the four angular momentums along the four directions zero. One can add these four angular momentum according to the following scheme:
\bea
|j_1,j_2,j_{12},j_{\bar 1},j_{\bar 2},j_{\bar 1\bar 2}=j_{12},j_{12\bar 1\bar 2}=0\rangle_x 
\eea
As, the abelian Gauss law implies $j_{\bar 1}|_x= j_1|_{x-e_1}$ and  $j_{\bar 2}|_x= j_2|_{x-e_2}$ following (\ref{u1gl}).
Hence, each site, one choice of such orthonormal basis at each site $x$ is \cite{pp}
\bea
|j_1,j_2,j_{12}\rangle_x
\eea
where, $j_1,j_2$ are the fluxes along $1~\&~ 2$ directions and $j_{12}$ is their added flux according to angular momentum addition scheme.
\label{secsqdyn}
Let us now characterize the state around a plaquette `abcd' as $$|j_{abcd}\rangle\equiv |j_1^a,j_2^a,j_{\bar 1}^a,j_{\bar 2}^a,j_{12}^a\rangle \times |j_1^b,j_2^b,j_{\bar 1}^b,j_{\bar 2}^b,j_{12}^b\rangle \times |j_1^c,j_2^c,j_{\bar 1}^c,j_{\bar 2}^c,j_{12}^c\rangle \times |j_1^d,j_2^d,j_{\bar 1}^d,j_{\bar 2}^d,j_{12}^d\rangle  $$
We further identify
\bea
j_1^a=j_{\bar 1}^b\equiv j_1 ~& ~j_2^b=j_{\bar 2}^c\equiv j_2 \\
j_{\bar 1}^c=j_{ 1}^d\equiv j_{\bar 1}  ~&~ j_{\bar 2}^d=j_{2}^a\equiv j_{\bar 2} 
\eea 
The dynamics of such states under the plaquette action are obtained as \cite{pp}
\bea
\langle \bar{j}_{abcd}|{\textrm Tr}U_{abcd}| {j}_{abcd} \rangle  & = &  
{\cal M}_{abcd}  
{\left \{ \begin{array}{cccc}
{j}_{1} &  \bar{j}_{1} & \frac{1}{2}  \\
\bar{j}_{\bar 2} & {j}_{\bar 2}  & {j}^a_{12}   \\ 
\end{array} \right\}}   
{\left\{ \begin{array}{cccc}
{j}^b_{12} & \bar{j}^b_{12} & \frac{1}{2}  \\
\bar{j}_{1} & j_{1} & j_{\bar 2}^b\\
\end{array} \right \}}  
{\left\{ \begin{array}{cccc}
{j}^b_{12} & \bar{j}^b_{12} & \frac{1}{2}  \\
\bar{j}_{2} & j_{2} & j_1^b\\
\end{array} \right \}} 
\nonumber  \\ \nonumber  \\ &&
\hspace{1cm} {\left \{ \begin{array}{cccc}
{j}_{\bar 1} &  \bar{j}_{\bar 1} & \frac{1}{2}  \\
\bar{j}_{2} & {j}_{2}  & {j}^c_{12}   \\ 
\end{array} \right\}}   
{\left\{ \begin{array}{cccc}
{j}^d_{12} & \bar{j}^d_{12} & \frac{1}{2}  \\
\bar{j}_{\bar 1} & j_{\bar 1} & j_2^d\\
\end{array} \right \}}  
{\left\{ \begin{array}{cccc}
{j}^d_{12} & \bar{j}^d_{12} & \frac{1}{2}  \\
\bar{j}_{\bar 2} & j_{\bar 2} & j_{\bar 1}^d \\
\end{array}\right\}}. ~~~~
\label{dyna1} 
\eea
In (\ref{dyna1}), ${\cal M}_{abcd} \equiv D_{abcd} N_{abcd} P_{abcd}$ factors are given by: 
\bea 
&&D_{abcd} =  \delta_{j^a_{\bar 1},\bar{j}^a_{\bar 1}} 
\delta_{j^a_{\bar 2},\bar{j}^a_{\bar 2}} 
\delta_{j^a_{12},\bar{j}^a_{12}}  
\delta_{j_1^b,\bar{j}_1^b} \delta_{j_{\bar 2}^b,\bar{j}_{\bar 2}^b}  
\delta_{j^c_{1},\bar{j}^c_{1}} \delta_{j^c_{2},\bar{j}^c_{2}} \delta_{j^c_{12},\bar{j}^c_{12}} 
\delta_{j_2^d,\bar{j}_2^d} \delta_{j_{\bar 1}^d,\bar{j}_{\bar 1}^d}, 
\nonumber \\ 
\label{cnf} 
&& N_{abcd} =
{\Pi}\left(j_1,\bar{j}_{1}, j_2,\bar{j_2},j_3,\bar{j_{\bar 1}},j_{\bar 2},\bar{j_{\bar 2}},j^b_{12},\bar{j}^b_{12},
j^d_{12},\bar{j}^d_{12}\right) 
\\
&& P_{abcd} =  - (-1)^{j_1+j_2+j_1^b+j_{\bar 2}^b} (-1)^{j_{\bar 1}+j_{\bar 2}+j_{\bar 1}^d+j_2^d}  
\triangle(\bar{j}_1,\bar{j}_{\bar 2},j_{12}^{a}) 
\triangle(\bar{j}_2,\bar{j}_{\bar 1},j_{12}^{c}) \nonumber \\
&&~~~~~~~~~~~~~~~~~
\triangle(\bar{j}^b_{12},{j}^b_{12},\frac{1}{2})  
\triangle(\bar{j}^d_{12},{j}^d_{12},\frac{1}{2}).   
\nonumber 
\eea

\begin{figure}[h]
\begin{center}
\includegraphics[width=0.8\textwidth,height=0.4\textwidth]
{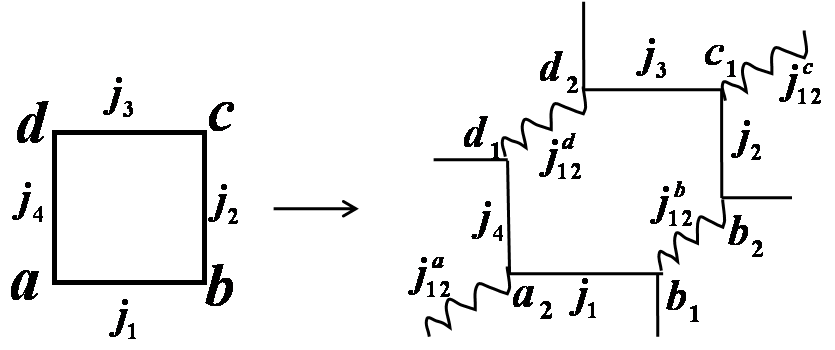}
\end{center}
\caption{Dynamics of loops on a square plaquette: orthonormal loop states are actually around the hexagon which fluctuate in plaquette dynamics. The hexagon arises as point splitting at each vertices, i.e $a\rightarrow a_1,a_2,~~b\rightarrow b_1,b_2,~~c\rightarrow c_1,c_2,~\& ~ d\rightarrow d_1,d_2,~~$} 
\label{sq2h}
\end{figure}

Note that, in (\ref{cnf}), $D_{abcd}$ describes the trivial $\delta$ functions over the angular 
momenta which do not change under the action of the plaquette operator, 
$N_{abcd}$ and $P_{abcd}$ give the corresponding numerical and the phase factors respectively. 
The multiplicity factors are:  $\Pi(x,y,...) \equiv \sqrt{(2x+1)(2y+1)...}$ and 
$\triangle(x,y,z)$ represent the phase factors given by: 
$\triangle(x,y,z)  \equiv  (-1)^{x+y+z}$.   
The $6j$ symbols in (\ref{dyna1}) yields \cite{varsh} a set of non-trivial delta functions given by:
$$\delta_{j_1,\bar j_1\pm \frac{1}{2}}\delta_{j_2,\bar j_2\pm \frac{1}{2}}\delta_{j_{\bar 1},\bar j_{\bar 1}\pm \frac{1}{2}}\delta_{j_{\bar 2},\bar j_{\bar 2}\pm \frac{1}{2}}\delta_{j_{12}^b,\bar j_{12}^b\pm \frac{1}{2}}\delta_{j_{12}^d,\bar j_{12}^d\pm \frac{1}{2}}$$
Now, looking at the expression for the dynamics of the theory we readily observe that, the four fluxes $j_1,j_2,j_{\bar 1},j_{\bar 2}$ flowing along the four sides of the plaquette do fluctuate by $\pm \frac{1}{2}$ units. Moreover, the plaquette action again changes  two intermediate fluxes, namely,  $j_{12}^b$ and $j_{12}^d$ in the same way as the four sides. We illustrate this fact in figure \ref{sq2h}.
At this point we get the motivation for formulating the gauge theory on hexagonal lattice and calculating the plaquette dynamics. 
\subsection{Evaluating $6j$ symbols}

The 6j symbols in (\ref{dyna1}) are of a special type, which containing one variable equal to $1/2$. There exists a simple prescription as given below, where one can convert these particular $6j$ symbols to those with one variable equal to $0$ as given below \cite{varsh}:
\bea
f_1{\left\{ \begin{array}{cccc}
a& b & \frac{1}{2}  \\
d & e &f\\
\end{array} \right \}} 
&=& f_2 {\left\{ \begin{array}{cccc}
a- \frac{1}{2}& b & 0  \\
d-\frac{1}{2} & e &f-\frac{1}{2}\\
\end{array} \right \}} +f_3 {\left\{ \begin{array}{cccc}
a- \frac{1}{2}& b & 0  \\
d-\frac{1}{2} & e &f+\frac{1}{2}\\
\end{array} \right \}} \nonumber \\
&&
+f_4 {\left\{ \begin{array}{cccc}
a- \frac{1}{2}& b & 0  \\
d+\frac{1}{2} & e &f-\frac{1}{2}\\
\end{array} \right \}} 
+f_5 {\left\{ \begin{array}{cccc}
a- \frac{1}{2}& b & 0  \\
d+\frac{1}{2} & e &f+\frac{1}{2}\\
\end{array} \right \}} 
\label{6j1}
\eea
with
\bea
f_1 &=& (2d+1)(2f+1)\sqrt{ (a+b+\frac{3}{2}) (a-b+ \frac{1}{2})} \nonumber \\
f_2 &=& -\sqrt{ (a+e+f+1)(a-e+f)(b+d+f+1)(-b+d+f)(d+e+\frac{3}{2})(\frac{1}{2}+d-e) } \nonumber \\
f_3 &=& -\sqrt{ (-a+e+f+1)(a+e-f)(b-d+f+1)(b+d-f)(d+e+\frac{3}{2})(\frac{1}{2}+d-e) } \nonumber \\
f_4 &=& -\sqrt{ (a+e+f+1)(a-e+f)(b+d-f+1)(b-d+f)(d+e+\frac{1}{2})(\frac{1}{2}-d+e) } \nonumber \\
f_5 &=& -\sqrt{ (-a+e+f+1)(a+e-f)(b+d+f+2)(-b+d+f)(d+e+\frac{1}{2})(\frac{1}{2}-d+e) } \nonumber 
\eea 
 
The resulting $6j$ symbols are easy to compute by the formula:
\bea
\label{6j2}
 {\left\{ \begin{array}{cccc}
\tilde a & \tilde  b & 0  \\
\tilde d &\tilde  e &\tilde f\\
\end{array} \right \}} = (-1)^{\tilde a+\tilde e+\tilde f} \frac{ \delta_{\tilde a,\tilde b}\delta_{\tilde d,\tilde e} }{\sqrt{(2\tilde a+1)(2\tilde d+1)}}
\eea


\begin{thebibliography}{99}
\bibitem{wil} K.~G.~Wilson,
  Phys.\ Rev.\ D {\bf 10}, 2445 (1974).
\bibitem{Creutz} M. Creutz. 1985. Quarks, Gluons and Lattices, Cambridge University Press; Heinz J. Rothe. 2003. World Scientific. Lattice Gauge Theories: An Introduction (Third Edition) 2005, John B. Kogut, Mikhail A. Stephanov The Phases of Quantum Chromodynamics: From Confinement to Extreme Environments, Cambridge University Press.
 \bibitem{kogut} J. Kogut, L. Susskind, 
 Phys. Rev. {\bf D 11} (1975) 395. 
\bibitem{pp} Manu Mathur, J. Phys. A: Math. Gen. {\bf 38} (2005) 10015; ;  Nucl. Phys.  B {\bf 779}, 32 (2007);
   Phys. Letts. {\bf B 640} (2006) 292-296;  R.~Anishetty, M.~Mathur and I.~Raychowdhury,
  J.\ Phys.\ A {\bf 43}, 035403 (2010)
  [arXiv:0909.2394 [hep-lat]]; I. Raychowdhury,  PhD thesis (2014).
  \bibitem{loops}R. Gambini. 2000. Jorge Pullin, Loops, Knots, Gauge Theories and Quantum Gravity (Cambridge University Press); Y.~Makeenko and A.~A.~Migdal,
  Nucl.\ Phys.\ B {\bf 188}, 269 (1981)
  [Sov.\ J.\ Nucl.\ Phys.\  {\bf 32}, 431 (1980)]
  [Yad.\ Fiz.\  {\bf 32}, 838 (1980)]; B.~Bruegmann,
  Phys.\ Rev.\ D {\bf 43}, 566 (1991); R.~Giles,
  Phys.\ Rev.\ D {\bf 24}, 2160 (1981); 
  W.~Furmanski and A.~Kolawa,
  Nucl.\ Phys.\ B {\bf 291}, 594 (1987); 
  R.~Gambini, L.~Leal and A.~Trias,
  Phys.\ Rev.\ D {\bf 39}, 3127 (1989);
  C.~Di Bartolo, R.~Gambini and L.~Leal,
  Phys.\ Rev.\ D {\bf 39}, 1756 (1989); R.~Loll,
  Nucl.\ Phys.\ B {\bf 368}, 121 (1992); A.~A.~Migdal,
  Phys.\ Rept.\  {\bf 102}, 199 (1983).
  \bibitem{mans} S.~Mandelstam,
  Phys.\ Rev.\ D {\bf 19}, 2391 (1979); R.~Loll,
  Nucl.\ Phys.\ B {\bf 400}, 126 (1993); N.~J.~Watson,
  Phys.\ Lett.\ B {\bf 323}, 385 (1994).
\bibitem{prd} R.~Anishetty and I.~Raychowdhury,
 Phys.\ Rev.\ D {\bf 90}, no. 11, 114503 (2014)
  [arXiv:1408.6331 [hep-lat]]; PoS LATTICE {\bf 2014}, 313 (2014)
  [arXiv:1411.3068 [hep-lat]].
  \bibitem{sreeraj} M.~Mathur and T.~P.~Sreeraj,
  Phys.\ Lett.\ B {\bf 749}, 137 (2015)
  [arXiv:1410.3318 [hep-lat]]; Phys.\ Rev.\ D {\bf 92}, no. 12, 125018 (2015)
  [arXiv:1509.04033 [hep-lat]]; 
  Phys.\ Rev.\ D {\bf 94}, no. 8, 085029 (2016)
  [arXiv:1604.00315 [hep-lat]].
\bibitem{new} R.~Anishetty and T.~P.~Sreeraj,
  arXiv:1802.06198 [hep-lat].
  \bibitem{qs} E.~Zohar, J.~I.~Cirac and B.~Reznik,
  Rept.\ Prog.\ Phys.\  {\bf 79}, no. 1, 014401 (2016)
  [arXiv:1503.02312 [quant-ph]]; M.~Dalmonte and S.~Montangero,
  Contemp.\ Phys.\  {\bf 57}, no. 3, 388 (2016)
  [arXiv:1602.03776 [cond-mat.quant-gas]].
  \bibitem{muller-ruhl} 
  V.~F.~Muller and W.~Ruhl,
  Nucl.\ Phys.\ B {\bf 230}, 49 (1984).
\bibitem{arisue} 
  H.~Arisue,
  Prog.\ Theor.\ Phys.\  {\bf 84}, 951 (1990).
\bibitem{Guo} S.~h.~Guo, J.~m.~Liu and W.~h.~Zheng,
  Phys.\ Rev.\ D {\bf 38}, 2591 (1988).
  \bibitem{mc} K.~Farakos, G.~Koutsoumbas and S.~Sarantakos,
  Phys.\ Lett.\ B {\bf 189}, 173 (1987).
  \bibitem{cluster} C.~J.~Hamer and A.~C.~Irving,
  Z.\ Phys.\ C {\bf 27}, 307 (1985).
 \bibitem{varsh} D. A. Varshalovich, A. N. Moskalev and V. K. Khersonskii. 1988. World Scientific. Quantum Theory of Angular Momentum.
 \end{thebibliography}
\end{document}